\documentclass[letter,apj]{emulateapj-rtx4}
\usepackage{graphicx}
\usepackage{enumerate}
\usepackage{amssymb, amsmath}
\usepackage{natbib}

\newcommand{\princeton}{1}
\newcommand{\goddard}{2}
\newcommand{\physmath}{3}
\newcommand{\fizeau}{4}
\newcommand{\mpia}{5}
\newcommand{\charleston}{6}
\newcommand{\subaru}{7}
\newcommand{\sternwarte}{8}
\newcommand{\naoj}{9}
\newcommand{\ifahawaii}{10}
\newcommand{\tokyo}{11}
\newcommand{\sokendai}{12}
\newcommand{\hiroshima}{13}
\newcommand{\madrid}{14}
\newcommand{\jpl}{15}
\newcommand{\sinica}{16}
\newcommand{\pannekoek}{17}
\newcommand{\sapporo}{18}
\newcommand{\oklahoma}{19}
\newcommand{\sendai}{20}

\begin{document}

\title{New Techniques for High-Contrast Imaging with ADI: the ACORNS-ADI SEEDS Data Reduction Pipeline}
\author{Timothy D.~Brandt\altaffilmark{\princeton}, Michael W.~McElwain\altaffilmark{\goddard}, Edwin L.~Turner\altaffilmark{\princeton,\physmath},
   L. Abe\altaffilmark{\fizeau},   
   W. Brandner\altaffilmark{\mpia},   
   J. Carson\altaffilmark{\charleston},    
   S. Egner\altaffilmark{\subaru},   
   M. Feldt\altaffilmark{\mpia},  
   T. Golota\altaffilmark{\subaru},   
   M. Goto\altaffilmark{\sternwarte},  
   C.~A. Grady\altaffilmark{\goddard}, 
   O. Guyon\altaffilmark{\subaru},  
   J. Hashimoto\altaffilmark{\naoj}, 
   Y. Hayano\altaffilmark{\subaru},  
   M. Hayashi\altaffilmark{\naoj},  
   S. Hayashi\altaffilmark{\subaru}, 
   T. Henning\altaffilmark{\mpia},   
   K.~W. Hodapp\altaffilmark{\ifahawaii},  
   M. Ishii\altaffilmark{\subaru}, 
   M. Iye\altaffilmark{\naoj},    
   M. Janson\altaffilmark{\princeton},   
   R. Kandori\altaffilmark{\naoj},  
   G.~R. Knapp\altaffilmark{\princeton},   
   T. Kudo\altaffilmark{\naoj},   
   N. Kusakabe\altaffilmark{\naoj},   
   M. Kuzuhara\altaffilmark{\naoj,\tokyo}, 
   J. Kwon\altaffilmark{\naoj,\sokendai},
   T. Matsuo\altaffilmark{\naoj},   
   S. Miyama\altaffilmark{\hiroshima},   
   J.-I. Morino\altaffilmark{\naoj},  
   A. Moro-Mart\'in\altaffilmark{\madrid},  
   T. Nishimura\altaffilmark{\subaru},
   T.-S. Pyo\altaffilmark{\subaru}, 
   E. Serabyn\altaffilmark{\jpl},   
   H. Suto\altaffilmark{\naoj},    
   R. Suzuki\altaffilmark{\naoj},   
   M. Takami\altaffilmark{\sinica},  
   N. Takato\altaffilmark{\subaru}, 
   H. Terada\altaffilmark{\subaru},  
   C. Thalmann\altaffilmark{\pannekoek},  
   D. Tomono\altaffilmark{\subaru},  
   M. Watanabe\altaffilmark{\sapporo},  
   J.~P. Wisniewski\altaffilmark{\oklahoma}, 
   T. Yamada\altaffilmark{\sendai},   
   H. Takami\altaffilmark{\subaru},  
   T. Usuda\altaffilmark{\subaru}, 
   M. Tamura\altaffilmark{\naoj}  
}

\altaffiltext{*}{Based on data collected at Subaru Telescope, which
   is operated by the National Astronomical Observatory of Japan.}

\altaffiltext{\princeton}{Department of Astrophysical Sciences, Princeton University, Princeton, USA.}
\altaffiltext{\goddard}{Goddard Space Flight Center, Greenbelt, USA.}
\altaffiltext{\physmath}{Institute for the Physics and Mathematics of the Universe, University
    of Tokyo, Japan.}
\altaffiltext{\fizeau}{Laboratoire Hippolyte Fizeau, Nice, France.}
\altaffiltext{\mpia}{Max Planck Institute for Astronomy, Heidelberg, Germany.}
\altaffiltext{\charleston}{College of Charleston, Charleston, South Carolina, USA.}
\altaffiltext{\subaru}{Subaru Telescope, Hilo, Hawai`i, USA.}
\altaffiltext{\sternwarte}{Universit\"ats-Sternwarte M\"unchen, Munich, Germany}
\altaffiltext{\naoj}{National Astronomical Observatory of Japan, Tokyo, Japan}
\altaffiltext{\ifahawaii}{Institute for Astronomy, University of Hawai`i, Hilo, Hawai`i, USA.}
\altaffiltext{\tokyo}{University of Tokyo, Tokyo, Japan.}
\altaffiltext{\sokendai}{Department of Astronomical Science, Graduate University for Advanced Studies, Tokyo, Japan}
\altaffiltext{\hiroshima}{Hiroshima University, Higashi-Hiroshima, Japan}
\altaffiltext{\madrid}{Department of Astrophysics, CAB - CSIC/INTA, Madrid, Spain.}
\altaffiltext{\jpl}{Jet Propulsion Laboratory, California Institute of Technology, Pasadena, CA, USA.}
\altaffiltext{\sinica}{Institute of Astronomy and Astrophysics, Academia Sinica, Taipei, Taiwan.}
\altaffiltext{\pannekoek}{Anton Pannekoek Astronomical Institute, 
    University of Amsterdam, Amsterdam, The Netherlands.}
\altaffiltext{\sapporo}{Department of Cosmosciences, Hokkaido University, Sapporo, Japan.}
\altaffiltext{\oklahoma}{HL Dodge Department of Physics and Astronomy, University of Oklahoma, Norman, OK, USA.}
\altaffiltext{\sendai}{Astronomical Institute, Tohoku University, Sendai, Japan}

\begin{abstract}

We describe Algorithms for Calibration, Optimized Registration, and Nulling the Star in Angular Differential Imaging (ACORNS-ADI), a new, parallelized software package to reduce high-contrast
imaging data, and its application to data from the SEEDS survey.  We
implement several new algorithms, including a method to register
saturated images, a trimmed mean for combining an image sequence that
reduces noise by up to $\sim$20\%, and a robust and computationally
fast method to compute the sensitivity of a high-contrast observation
everywhere on the field-of-view without introducing artificial sources.  We also
include a description of image processing steps to remove electronic artifacts specific to Hawaii2-RG
detectors like the one used for SEEDS, and a detailed
analysis of the Locally Optimized Combination of Images (LOCI)
algorithm commonly used to reduce high-contrast imaging data.  ACORNS-ADI is written in python.  It is efficient and open-source, and includes several
optional features which may improve performance on data from other
instruments.  ACORNS-ADI requires minimal modification to reduce data from instruments other than HiCIAO.
It is freely available for download at \verb|www.github.com/t-brandt/acorns-adi| under a BSD license.
\end{abstract}

\section{Introduction}

Since 1992, more than 700 confirmed exoplanets and 2000 additional candidates have been discovered\footnote{See http://www.exoplanet.eu/}.  Ground-based surveys have confirmed hundreds of exoplanets by measuring the periodic radial velocity shifts they induce in their host stars (e.g., \citealt{Vogt+Marcy+Butler+etal_2000, Queloz+Mayor+Weber+etal_2000, Tinney+Butler+Marcy+etal_2001, Mayor+Pepe+Queloz+etal_2003}) or by measuring photometric variations as they transit their host stars (e.g., \citealt{Alonso+Brown+Torres+etal_2004, Bakos+Noyes+Kovacs+etal_2004, McCullough+Stys+Valenti+etal_2005, Pollacco+Skillen+CollierCameron+etal_2006, Charbonneau+Berta+Irwin+etal_2009}).  In space, NASA's {\it Kepler} satellite \citep{Borucki+Koch+Basri+etal_2010} has identified more than 2000 candidate transiting exoplanets.  These indirect methods are sensitive to short-period exoplanets: the magnitude of a radial velocity signal and the probability of a transit both decrease with separation.  These methods also generally require observations over several orbital periods, making them impractical for detecting exoplanets with periods of more than a few years.  

Direct imaging surveys, made possible by advances in adaptive optics, infrared detectors, and image processing algorithms, are now complementing transit and radial velocity surveys, identifying giant exoplanets tens of astronomical units (AU) from their host stars.  Ground-based high-contrast imaging surveys have shown that these giant exoplanets are rare (e.g., \citealt{Biller+Close+Maciadri+etal_2007, Lafreniere+Doyon+Marois+etal_2007}), and are beginning to constrain models of exoplanet and exoplanetary system formation and evolution \citep{Janson+Bonavita+Klahr+etal_2012}.  

Large-scale direct-imaging surveys rely on sophisticated image processing to search for faint companions around bright stars.  In
addition to the usual bias, flat-field, and distortion corrections, these surveys must model and subtract the stellar point-spread function (PSF).  Most surveys use Angular Differential Imaging (ADI) to make this task easier.  As the Earth rotates, the orientation of the field-of-view (FOV) of an altitude-azimuth telescope (and thus of the PSF on the detector) changes relative to the celestial north.  Features of the PSF due to the instrument, such as telescope spiders and the diffraction pattern, appear to rotate relative to any faint companion.

The first algorithms to take advantage of ADI used simple techniques to model the PSF, like taking the median of a sequence of exposures \citep{Marois+Lafreniere+Doyon+etal_2006}.  More recently, algorithms like the Locally Optimized Combination of Images \citep[LOCI,~][]{Lafreniere+Marois+Doyon+etal_2007} model the PSF locally, while principal component analysis (PCA)-based techniques \citep{Soummer+Pueyo+Larkin_2012,Amara+Quanz_2012} model it globally.  These more sophisticated algorithms can offer a factor of $\sim$2 or more improvement in sensitivity over simple ADI reductions.

In this paper, we present Algorithms for Calibration, Optimized Registration, and Nulling the Star in Angular Differential Imaging (ACORNS-ADI), a software package to analyze ADI data for the SEEDS survey (see \citealt{Tamura_2009}), a five-year direct imaging survey using the HiCIAO instrument \citep{Hayano+Takami+Guyon+etal_2008} on the Subaru Telescope.  
We discuss each non-trivial step of the reduction process, from the bias and flat-field corrections to PSF modeling to the final sensitivity analysis.  ACORNS-ADI is parallelized, open-source, and freely available for download at \verb|www.github.com/t-brandt/acorns-adi| under a BSD license.

\section{ADI Data Reduction in SEEDS} \label{sec:adioverview}

In order to take advantage of the field rotation in ADI, a series of short exposures is taken, with minimal field rotation during each individual exposure.  The central star is usually allowed to saturate in order to increase the observing efficiency and limit the amount of read noise for a given number of companion photons.  A typical high-contrast ADI dataset thus consists of a series of short, sequential exposures with a saturated central star.  The data reduction process searches for point sources in the image sequence.  
While it is also possible to analyze extended structures, like disks, using ADI reduction techniques \citep[e.g.][]{Liu_2004, Thalmann+Janson+Buenzli+etal_2011}, it is far more difficult to interpret the final processed images (see Section \ref{sec:medpsf}).  A detailed discussion of extended sources in ADI is beyond the scope of this paper.

A reduction of a high-contrast ADI image sequence proceeds in several
steps:
\begin{enumerate}
\item Correct for the bias, flat-field each image;
\item Interpolate over hot pixels;
\item Correct the image distortion (if necessary);
\item Register the frames (if necessary);
\item Model and subtract the PSF of the central star;
\item Rotate each frame to align it to the celestial north;
\item Combine the sequence of images;
\item Search for point sources; and
\item Produce a sensitivity map.
\end{enumerate}
Each step in the reduction impacts the sensitivity of the final
combined image.  Optimizing and characterizing this sensitivity is
critical for understanding the incidence and properties of substellar companions.

Some of the steps listed above, such as the distortion correction (once the
distortion map is known!) and the rotation to a common frame, are
trivial.  Others are surprisingly difficult, or can be optimized to
give significant improvements in sensitivity.  For example, a
one-pixel root-mean-square scatter in the image registration can
degrade sensitivity by $\sim$20\% (Section \ref{sec:loci}), as can the
use of the median intensity, rather than a trimmed mean, to combine a sequence of images (Section
\ref{sec:mean_median}).

\begin{figure*}
\centering\includegraphics[width=0.9\linewidth]{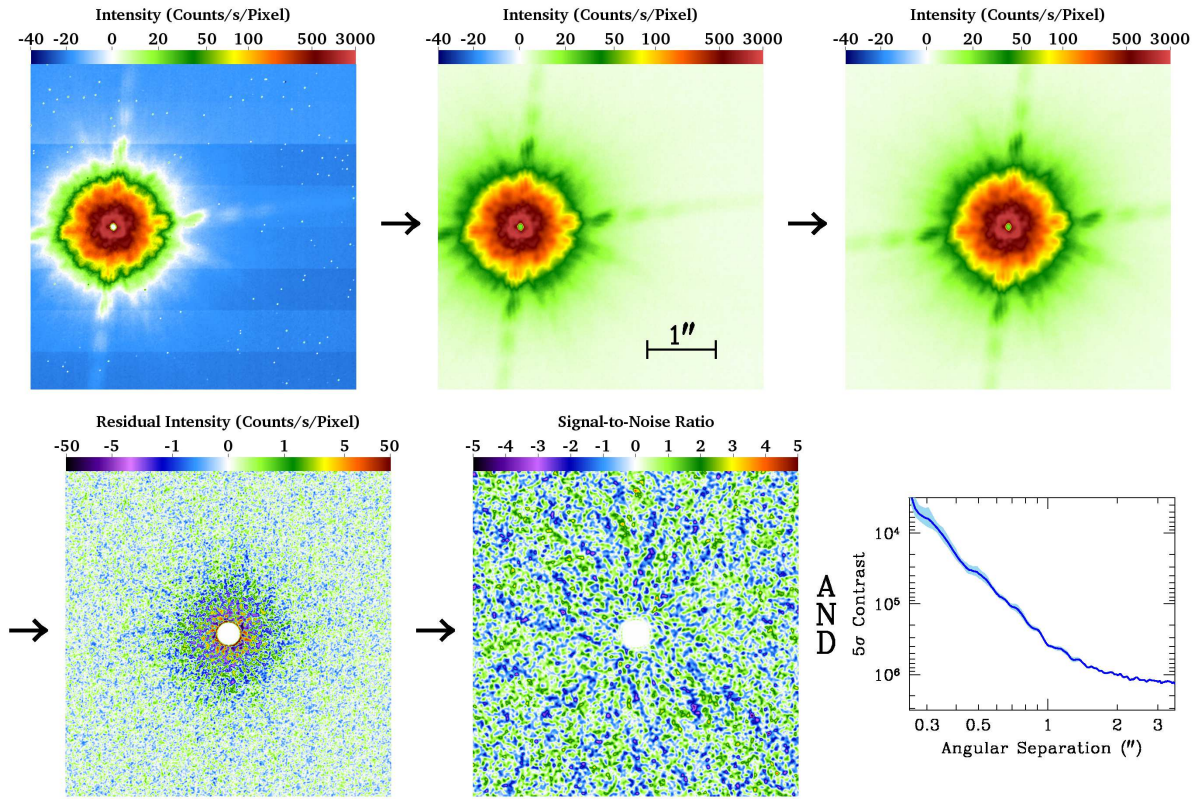}
\caption{A step-by-step depiction of the ACORNS-ADI data reduction process,
  as discussed in Section \ref{sec:adioverview}.  The first image
  shows the central $5'' \times 5''$ of a $20''\times 20''$ frame of raw data.  The second frame has been bias-corrected, flat-fielded, and had its hot pixels masked
  (Sections \ref{sec:destripe} and \ref{sec:flat_mask}), while the
  third frame has been corrected for field distortion (Section
  \ref{sec:distortion}) and registered to the PSF centroid (Section
  \ref{sec:center}).  The first frame on the second row shows the
  residuals in a single frame after applying the LOCI algorithm
  (Section \ref{sec:loci}, Appendix \ref{app:locisub}), while the next
  frame shows the combined image from an ADI sequence (Section
  \ref{sec:mean_median}, Appendix \ref{app:meanmedstats}) convolved
  with a circular $0.\!\!''05$ aperture (Section \ref{sec:filter}) and
  normalized by the standard deviation at each radius.  The final
  frame shows a radial profile, with shading to represent azimuthal scatter,
  of the final sensitivity map (Section \ref{sec:sensitivity}, Appendix
  \ref{app:locisub}).}
\label{fig:pipeline_stepbystep}
\end{figure*}

Figure \ref{fig:pipeline_stepbystep} shows a sample SEEDS dataset
through the above sequence of steps as processed by ACORNS-ADI.  The first frame shows the central $5'' \times 5''$ of a $20'' \times 20''$ sample
raw image, while the second frame shows the effect of correcting for
the bias (Section \ref{sec:destripe}), flat-fielding, and hot pixel
masking (Section \ref{sec:flat_mask}).  The third frame shows the same
image after correcting for field distortion (Section
\ref{sec:distortion}) and registering to the PSF centroid (Section
\ref{sec:center}).  The first frame on the second line shows the
residuals after subtracting the stellar PSF using the LOCI algorithm
(Sections \ref{sec:loci} and \ref{sec:loci_calib}, Appendix \ref{app:locisub}); the next frame
shows the combined image from an ADI sequence (Section
\ref{sec:mean_median}, Appendix \ref{app:meanmedstats}) convolved with
a circular $0.\!\!''05$ aperture (Section \ref{sec:filter}) and normalized by
the radial profile of its standard deviation.  Finally, we show a radial
profile of the dataset's sensitivity to point sources (Section \ref{sec:sensitivity}, Appendix \ref{app:locisub}).

In the following sections, we describe each non-trivial step listed
above.  Some steps, like the subtraction of the stellar PSF and the
combination of an image sequence, apply generally to data from any
survey, while other steps, like the distortion correction and bias
correction, have features specific to SEEDS data.  Our discussions of
two of these steps, the computation of a sensitivity map and the
statistics of a combination of images, contain calculations that we
relegate to appendices.

\section{Calibration}

The first step in the data reduction is calibration:
finding the count rate corresponding to zero intensity, flat-fielding,
and applying a distortion correction.  The zero point
correction and flat-fielding are complex and interrelated for SEEDS data, and we
handle them with a single routine.  We then apply a distortion correction to
the intensity-calibrated data.  Because an ADI sequence consists of many short exposures, its sensitivity far from the central star can be limited by calibration uncertainties and read noise.  Typical SEEDS observations are read noise limited at separations $\gtrsim$$2''$.

The algorithms and discussions in this section are mostly specific to
data from a 2048$\times$2048 pixel Hawaii2-RG (H2RG) detector \citep{Blank+Anglin+Beletic+etal_2011}.  These HgCdTe detectors are
becoming more common on major instruments and telescopes.  In addition
to HiCIAO on Subaru, Calar Alto, CFHT, VLT, IRTF, Keck, and SALT all use H2RG detectors.  Future space-based missions such as {\it JWST}, {\it JDEM}, {\it EUCLID}, and Prime Focus Spectrograph and CHARIS, the next-generation spectrograph and camera for Subaru, will
all contain H2RG or similar, but larger, 4096$\times$4096 pixel H4RG arrays.  To reduce data from another
detector, the user would need to supply his or her own flat frame and 
bad pixel mask, and configure ACORNS-ADI to skip the steps listed below.
This can be done during ACORNS' interactive configuration without any
modification of the source code.

\subsection{Removing the Bias} \label{sec:destripe}

As configured for SEEDS, HiCIAO's H2RG detector reads out data in 32
channels at a pixel rate of 100 kHz with correlated double sampling (CDS).  Reading out all $2048 \times
2048$ pixels thus takes about 1.3 seconds.  H2RG detectors also feature a non-destructive read mode, so that up-the-ramp sampling could be implemented in the future on longer exposures.  
As discussed by
\cite{Moseley+Arendt+Fixsen+etal_2010}, each of the 32 readout
channels has its own reference voltage, which appears as a bias---a non-zero count level corresponding to zero intensity---in raw HiCIAO data.  Superimposed on this is a time-varying
reference voltage which is largely shared between the 32 channels.  To
calibrate the count level corresponding to zero intensity, we therefore need to fit for 32 stable
reference voltages and at least one function.  The H2RG detector
provides four rows of pixels at each detector edge, for a total of
32,704 reference pixels (the `R' in H2RG).  These pixels are not light-sensitive, but
are subject to the same reference voltages as the rest of the array,
and can be used to estimate the pixel-by-pixel bias.  For some
observations, SEEDS has taken data with the guide window
capability (the `G' in H2RG), reading out only a subarray of the detector.  In this mode,
there is a single readout channel (and reference voltage) and there
are no reference pixels.

For SEEDS data taken in the normal 32-channel readout mode, the 512
reference pixels at the ends of the channel are sufficient to
determine the stable reference voltages.  The H2RG read noise is very
nearly Gaussian, so we use a sigma-reject technique to calculate their
offset from zero.  ACORNS-ADI iteratively rejects 3$\sigma$
outliers, takes the mean of the remaining reference pixels, and
subtracts this value from all of the pixels in each readout channel.
This calibration is good to approximately the read noise over
$\sqrt{512}$ (the number of reference pixels), or about 1 e$^-$ per frame.  The first row of Table
\ref{tab:destripe} shows the read noise as measured in a series of 30
dark frames taken in December 2010 with no bias corrections.
The difference between the root variance within a readout channel (27
e$^-$) and over the entire array (52 e$^-$) is almost completely
removed by a mean subtraction using only the 512 reference pixels, as
shown in the second row of Table \ref{tab:destripe}.

The time-varying reference voltage, largely shared among the 32
readout channels, is more difficult to subtract.
\cite{Moseley+Arendt+Fixsen+etal_2010} describe two techniques.  One
of these relies on reference pixels interspersed throughout the
detector array, which would be difficult to implement in an imaging survey like
SEEDS.  The other technique saves
the reference voltage in place of one of the 32 channels of science pixels.
Appropriate weighting of the Fourier components of this reference
voltage then provides a good estimate of the bias to be subtracted
from each readout channel.  

The H2RG detector has $1/f$ noise that
extends from the frame rate, $\sim$1 Hz, to a knee at $\sim$3 kHz, where
the noise becomes uncorrelated and 
the optimal weighting of the Fourier components falls to zero
\citep{Moseley+Arendt+Fixsen+etal_2010}.  
SEEDS does not currently save the
reference voltage, which would require the sacrifice of $1/32$ of the
FOV; we have therefore estimated the possible improvement in read
noise by applying a high-pass filter, removing all read noise with a frequency $\lesssim 3$ kHz from
each channel.  Removing all low-frequency noise reduces the total read noise by about
10\%, from 27 e$^-$ to 25 e$^-$ (fourth row of Table
\ref{tab:destripe}).

Without the reference voltage, ACORNS-ADI implements two
techniques to estimate and subtract the reference voltage; the user
must select which to use.  The first technique uses the reference
pixels at the edges of channels 1 and 32, which provide eight
measurements of the reference voltage every 64 pixels.  We first remove 
outliers with sigma-rejection, and then subtract the convolution
of the time series of reference pixels with a masked Gaussian.  
The Gaussian is normalized to unit area, and is zero where no reference
voltage is available.  We choose its width to optimize the reference 
subtraction as follows.

While the H2RG detector has $1/f$ noise that
extends up to a knee at $\sim$3 kHz,
sparse sampling (and read noise) limit our ability to measure the
high frequency noise.  As shown in Table
\ref{tab:destripe}, a perfect suppression of the noise up to $\sim$3
kHz would decrease the read noise by $\sim$10\%, or the variance by
just under 20\%.  However, a poor estimate of the reference voltage
can increase the noise.  We can then estimate our fractional
suppression of the read noise as
\begin{equation}
\frac{0.2}{\ln 3000} 
\int_{1\,\rm Hz}^{\nu_{\rm max}} \frac{d\nu}{\nu} - \frac{1}{N} ~,
\label{eq:biassub}
\end{equation}
where $N$ is the effective number of pixels used to estimate the
reference voltage at each point, $\nu_{\rm max} \propto N^{-1}$, and $\ln 3000$ is the value of the integral with $\nu_{\rm max} = 3$ kHz.
Maximizing Eq.~\eqref{eq:biassub}, we find that $N \sim 40$.  Since
only one in eight pixels has a corresponding reference pixel, this is
equivalent to an effective smoothing length of $\sim$300 science
pixels.  We therefore cannot suppress noise of $\gtrsim$300 Hz, and
achieve a $\sim$5\% reduction in read noise (third row of Table
\ref{tab:destripe}) rather than a $\sim$10\% reduction (fourth row).
Like \cite{Moseley+Arendt+Fixsen+etal_2010}, we optimally suppress
low-frequency read noise by scaling the reference signal; we find a
best-fit scaling of 0.87.

ACORNS-ADI also implements a technique to remove correlated
read noise using some or all of the
science pixels.  To give an accurate estimate of the bias, these pixels must be
uniformly illuminated.  SEEDS generally observes a saturated central
source whose seeing halo extends out to several hundred pixels in
radius (several arcseconds); pixels beyond this halo can be used to better estimate the
high frequency components of the reference voltage.  ACORNS-ADI
estimates the (uniform) illumination using the difference between the
reference and science pixels, and takes the median of each set of
up to 32 simultaneous readouts.  Using all of the science pixels,
this procedure reduces the read noise by $\sim$10\% over the perfect
subtraction of all read noise of frequency $\lesssim$3 kHz (last line
of Table \ref{tab:destripe}), much more than the $\sim$2\% expected
from self-subtraction.  The result is nearly as good when using only half
the science pixels, those at least 800 pixels from the center (fifth line of
Table \ref{tab:destripe}).  This level, $\sim$23 e$^-$, indicates the
read noise that is not shared between readout channels.  We recommend using 
unilluminated science pixels for read noise suppression if they are 
available.  We do not recommend a bias correction using only reference 
pixels unless the field is crowded.

\begin{deluxetable}{lcr}
\tablewidth{0pt}
\tablecaption{Average Residual Read Noise After Bias Correction
  in 30 Dark Frames}
\tablehead{
  \colhead{Zero Point Method} &
  \multicolumn{2}{c}{Average Read Noise (e$^- / {\rm pixel}$)} \\
  \colhead{} &
  \colhead{Single Channel} &
  \colhead{Entire FOV}
}

\startdata

None & $27.2$ & $51.8$ \\
One Voltage per Channel & $27.2$ & $27.4$ \\
Reference Pixels, $\nu <300$ Hz & $26.1$ & $26.2$ \\
All Pixels, $\nu <3$ kHz & $25.0$ & $25.1$ \\
Half of the Pixels, Median & $23.4$ & $23.8$ \\
All Pixels, Median & $22.3$ & $22.5$ 
\enddata
\label{tab:destripe}
\end{deluxetable}

\subsection{Flat-Fielding and Hot Pixel Masking} \label{sec:flat_mask}

For a typical exposure of $\sim$1 to 10 seconds, the dark current is
$\sim$0.05 to 0.5 e$^-$, a factor of $\sim$50 to 500 lower
than the read noise.  Pixel-to-pixel variations in the dark current
will be still lower.  A perfect suppression of the dark current would
reduce the read noise by $\lesssim$0.01\%; we therefore do not attempt
to correct for it, but treat the dark current as part of the
background.

HiCIAO shares the problem of ``hot,'' indium-contaminated pixels with
other H2RG detectors.  As a result, $\sim$2\% of HiCIAO's pixels
are unusable.  These hot pixels are stable from night-to-night, though
because the detector slowly degrades with time, new hot pixel maps
must be made periodically.  ACORNS-ADI corrects these
pixels by taking the median of all uncontaminated pixels in a $5
\times 5$ box centered on the hot pixel.  Because the data are oversampled, with a PSF core typically being $\sim$6 pixels in diameter, 
this will not significantly bias the intensity.  We mask hot pixels
throughout the bias calculation described above.

Because of the short exposure times in most SEEDS images, cosmic rays
are rarely a problem.  We do implement an algorithm to reject
isolated discrepant pixels or clusters of pixels similar to the one
used by \cite{Lafreniere+Marois+Doyon+etal_2007}.  We aggressively
smooth the image with a large median filter, identify pixels that
are above a certain threshold in the difference between the original
and the smoothed maps, and mask them.  

Flat-field images are stable to $\sim$2\% from month to month, and are
even more stable from night to night.  We therefore construct one
flat-field for each observing run from nine dithered dome flats.  We
correct the bias of each frame with the method described in
Section \ref{sec:destripe} using only the reference pixels, and then
median-combine the nine dithered images.  ACORNS-ADI divides
each science frame by this master flat after performing the bias
correction.  We do not attempt to correct for the detector's non-linearity 
as it approaches saturation.

\subsection{Distortion Correction} \label{sec:distortion}

The field distortion is the difference between positions on the
detector and positions on-sky.  We model the distortion using a
third-order two-dimensional (2D) polynomial in pixel cordinates relative to the center
of the FOV; the distortion at the image center is thus zero
by definition.  We use {\it Hubble Space Telescope} images of the
globular clusters M5 and M15 for our reference images, extracting
stellar positions using SExtractor \citep{Bertin+Arnouts_2010}.  We
use the same tool to extract stellar positions in HiCIAO images.  We
then fit for the polynomial coefficients, using Markov Chain Monte
Carlo \citep{Press+Teukolsky+Vetterling+etal_2007} to minimize the
difference between {\it Hubble} positions and corrected HiCIAO
positions.  This technique gives best-fit values and errors on all
parameters.  On a good night, we measure the horizontal and vertical
pixel scales to 0.01\%, and the direction of true north to a precision
of $\sim$$0.005^\circ$; poor conditions degrade our precision by a factor of up to $\sim$5.  Finally, we use bilinear interpolation to
transform our image to the new coordinate system, which has a pixel
scale of 9.5 mas, similar to the pixel scale of the raw data.  The distortion correction is described in more
detail by \citep{Suzuki+Kudo+Hashimoto+etal_2010}.

Our corrected images appear to be free of systematics, with the
distribution of offsets between {\it Hubble} and HiCIAO positions
being Gaussian in both the horizontal and vertical coordinates, and
random over the FOV.  The orientation of true north has been stable to $\sim$0.03$^\circ$
since the SEEDS survey began in 2009, but the plate scale has changed
by up to 2\% due to small changes in the optical setup and a new
camera lens which was installed in April of 2011.  Apart from changes
in the overall plate scale, the field distortion has been extremely stable
over the duration of the SEEDS survey.  It is dominated by a $\sim$3\%
difference between the horizontal and vertical pixel scales and a
$\sim$0.3$^\circ$ offset between the vertical axis on the detector and
the celestial north.  

\section{Image Registration} \label{sec:center}

Image registration is important both to optimize the subtraction of
the stellar PSF and to maximize companion flux in the final
de-rotated, co-added image.  Unfortunately, it is difficult to
centroid saturated stars, and most SEEDS data have no bright but
unsaturated stars in the $20'' \times 20''$ FOV.  Here, we present a
new algorithm to register isolated, saturated HiCIAO PSFs.  This
algorithm performs well, with a residual scatter of $\sim$1--2 mas (0.1--0.2 pixels) for
observations made in good conditions, and could easily be applied to
data from other instruments.

\subsection{Developing a PSF Template}

\begin{figure}
\includegraphics[width=\linewidth]{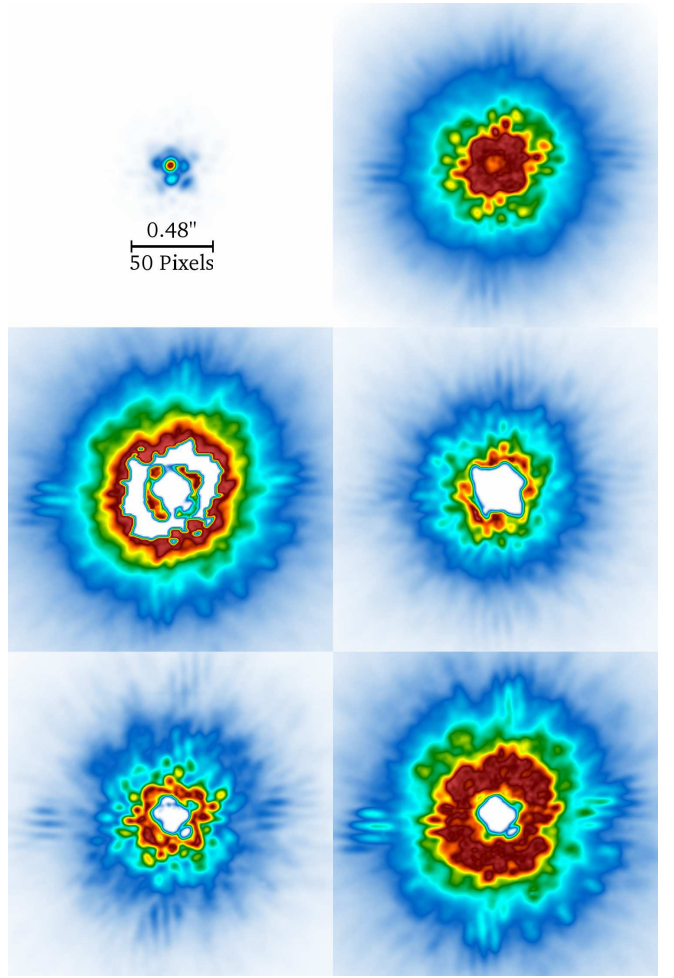}
\caption{Sample HiCIAO PSFs.  The top-left panel is unsaturated and
  includes the scale, while the other panels each represent a typical 
  PSF of a different object observed as part of the SEEDS survey.  The color
  scale is linear with white representing zero; because HiCIAO reads
  each pixel twice, areas that saturate rapidly can appear white.
  Using the algorithm described in Section \ref{sec:center}, we can
  centroid each PSF to an accuracy of better than 0.5 pixels.  For the
  better-behaved PSFs, like that shown the bottom left, our accuracy
  improves to $\sim$0.1--0.2 pixels, or $\sim$1--2 mas. }
\label{fig:sample_psfs}
\end{figure}

To accurately centroid images, we need to build a model of the HiCIAO
PSF.  Unfortunately, HiCIAO's adaptive optics (AO) system must be re-tuned at the
beginning of each observation, and its PSF varies accordingly.  Figure
\ref{fig:sample_psfs} shows this variation in a representative sample
of SEEDS images.  The top-left PSF is unsaturated, while the other panels each show a typical PSF of a different ADI sequence.  The PSF varies strongly with atmospheric conditions and the performance of AO188, HiCIAO's 188-actuator
AO system \citep{Hayano+Takami+Guyon+etal_2008}, which itself depends on stellar brightness.  To capture this variation, we proceed empirically, ultimately
building a set of three PSF templates from nearly 3000 individual
exposures of saturated stars observed as part of the SEEDS survey.

We extract three templates from 3000 images by iteratively registering
the frames and performing PCA.  We
initially centroid the frames by fitting Moffat profiles.  We manually
remove outliers---some of which result from bad data, others from the
failure of the algorithm---and use a scaled, unsaturated PSF to flag
saturated pixels.  We then rescale the data to a common flux and
perform PCA on the sequence of images.  We use weighted PCA,
estimating the noise at each unsaturated pixel as the sum of read
noise and photon noise, and ignore saturated pixels.

We now use the mean PSF and the first two principal components from
this first pass to refine our centroids.  We first re-estimate the
centroid of each image by flagging saturated pixels (those with at
least 70\% of the maximum intensity on the detector) and centroiding the
greatest concentration of such pixels.  We then compute the
root-mean-square distance $r_{\rm rms}$ of the saturated pixels from
the provisional center, and mask all pixels within $1.5 r_{\rm rms}$.
We model the variance of the intensity at each remaining pixel as shot
noise plus read noise, and fit the frame with a linear combination of
the mean PSF and the first two principal components by minimizing
\begin{equation}
\chi^2 = \sum_{{\rm pix}~i} \frac{1}{\sigma_i^2} 
\left( I_i - \sum_{{\rm tmpl}~j} \alpha_j T_{ij} \right)^2 ,
\end{equation}
where $j$ indexes the templates $T$ ($T_0$ represents the mean PSF),
$i$ indexes the pixels, and $\alpha_j$ are free parameters.  We then
shift the template PSF by an integer number of pixels relative to the image (thereby avoiding any artifacts from interpolation), and compute the best-fit
$\chi^2$ as a function of positional offset.  Finally, we centroid the
$\chi^2$ map by fitting a two-dimensional quadratic, and take the
location of this peak to be the centroid of the image.  We then
register all of the frames, scale them to a common flux, mask
saturated pixels, weight the other pixels, and perform PCA.  We repeat
this entire sequence of steps one more time to build our final set of
PSF templates, which we take to be the mean PSF and the first two
principal components.

\subsection{Centroiding Saturated Frames}

We register each sequence of saturated frames using the same method
described in the previous section.  The algorithm initially proceeds
in four steps:
\begin{enumerate}
\item Flag saturated pixels, centroid the greatest concentration of
  such pixels to compute a provisional center;
\item Mask pixels near the provisional center, weight all other pixels
  by photon and read noise;
\item Fit the PSF templates using $\chi^2$ at an integer-pixel grid of offsets; and
\item Centroid the map of $\chi^2$ merit statistics.
\end{enumerate}
We then recenter each frame and compute the average PSF of the ADI
sequence.  As Figure \ref{fig:sample_psfs} shows, the average PSF in a
given sequence of images can be significantly asymmetric.  We
therefore compute relative centroids according to the procedure just
described, and separately determine the centroid of the average PSF.
We do not re-compute a separate PSF model for each ADI sequence.  Doing so
offers no improvement in performance, and appears to introduce slight
interpolation artifacts.

\begin{figure}
\centering\includegraphics[width=0.9\linewidth]{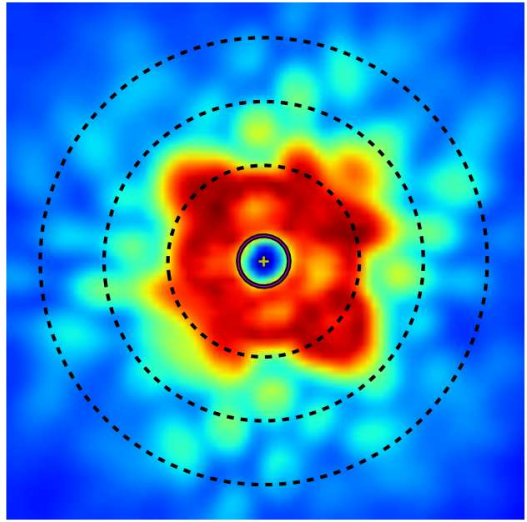}
\caption{Sample average HiCIAO PSF from an ADI sequence.  The user
  interactively chooses the absolute centroid in the average frame;
  the resulting offset is applied to each frame.  The yellow cross shows
  the center of the image while the innermost circle has a radius of 4
  HiCIAO pixels, or about 38 mas; the image is interpolated between points and therefore appears smooth.  The absolute centroids in a typical
  sequence may be estimated to an accuracy of $\lesssim$0.5 pixels
  $\approx$ 5 mas; the relative centroids, as discussed in Section
  \ref{sec:centroid_performance}, are better.}
\label{fig:sample_centroid}
\end{figure}

A peculiar feature of HiCIAO makes it possible to visually estimate
the centroid of a saturated frame.  The HgCdTe H2RG chip is reset
pixel-by-pixel and then read out twice; the interface computer records
the difference between the two readings.  As a result, pixels that
saturate rapidly tend to show no difference between the two readings,
and hence, are recorded as having zero intensity.  We therefore allow
the user to interactively select the absolute centroid of the average
of an ADI sequence.  Figure \ref{fig:sample_centroid} shows a sample
centroid verification image; the innermost circle has a radius of 4
HiCIAO pixels, or about 38 mas.  As Figure \ref{fig:sample_centroid}
indicates, it is usually possible to absolutely centroid a sequence of
saturated images to an accuracy of at least $\sim$0.5 pixels, around 5
mas.  We then add this user-determined offset to each frame's
centroid.  In this way, ACORNS-ADI determines the relative centroids
automatically, while the user selects the absolute centroid, in the
form of a single offset for the sequence of images.

Unfortunately, it is difficult to independently verify the accuracy of 
the absolute centroid.  The SEEDS survey does have a few image sequences
with a bright, unsaturated star, and those taken under favorable conditions
confirm an accuracy of $\lesssim$0.5 pixels.  Image sequences with poor 
AO performance appear to have larger errors of $\sim$1--2 pixels.  These errors remain impossible to 
verify on most sequences.  We recommend that the user choose a 
conservative error $\sigma_{\rm cen}$, and scale the sensitivity by the factor
\begin{equation}
\left( 1 + \frac{\sigma_\phi \sigma_{\rm cen}}{R_{\rm PSF}} \right)^{-1},
\end{equation}
where $\sigma_\phi$ is the standard deviation in parallatic angle and
$R_{\rm PSF}$ is the diameter of the PSF core.  Any error in the absolute
centroid will smear out a companion by a fractional amount roughly proportional
to the field rotation, and inversely proportional to the size of the PSF.

\subsection{Performance of the Centroiding Algorithm} \label{sec:centroid_performance}

We measure the performance of our new centroiding algorithm using ADI
image sequences of single stars.  The performance consists of the
centroids' relative accuracy, which is completely determined by ACORNS-ADI, and their absolute accuracy, which is determined by the user
and is much more difficult to assess.  The accuracy of the relative
centroids affects the quality of the data reduction, but not its
astrometry.  Poor image registration will smear out speckles and point
sources but without introducing systematic offsets.  The astrometric
error in the reduced data is thus equal to the error in the absolute
centroid, and must be estimated by the user.  This varies from dataset
to dataset but for data with good AO performance, is generally $\lesssim$0.5 pixels, $\sim$5 mas.

We can assign upper limits to the errors in image registration using
the scatter in fitted centroid positions.  This scatter will be due
both to tracking errors and PSF fitting errors, which we assume to be
uncorrelated.  We further assume that tracking errors dominate the
slow drift in the PSF position.  Before the installation of an
atmospheric dispersion corrector \citep[ADC,][]{Egner+Ikeda+Watanabe+etal_2010} in AO188,
the PSF would drift over the
course of an observation due to differential atmospheric refraction
between the visible, in which the AO system guides, and the
near-infrared in which HiCIAO observes.  This effect was mostly,
though not entirely, eliminated by the installation of the ADC.  More
recent observations at high airmass \citep{Thalmann+Janson+Buenzli+etal_2011} indicate
that these slow drifts occur at the level of at most a few pixels
(tens of mas) over the course of an ADI sequence.

As we wish to measure only the frame-to-frame fluctuations in the
fitted centroids, we fit and subtract a low-order polynomial from the
centroid positions in each ADI sequence.  An alternative approach, measuring
the positional difference between successive frames and dividing by $\sqrt{2}$,
gives nearly identical results.  We then compute the rms scatter of
the residual centroid positions, excluding the most discrepant 1\% of
points to remove outliers.  For a true Gaussian distribution, this
outlier exclusion reduces the variance by 10\%; we correct our
measured variances for this effect.  The worst datasets, those in
which the PSF varies most and positional jitter is most likely to be
real, show a root-mean-square (rms) scatter of $\sim$0.3 pixels (3
mas) in both the horizontal and vertical directions, or $\sim$0.4
pixels overall.  Most of the data are much better; 12 of the 21 ADI
sequences we used to build the template PSFs showed overall residual
rms scatters of less than 0.2 pixels (2 mas), and 17 of 21 had rms scatters
of less than 0.3 pixels (3 mas).  

Given the variation and asymmetries in HiCIAO's PSF, we allow the user
to interactively determine the absolute centroid of an image sequence.
However, the algorithm presented above can register a series of images
to a typical precision of $\sim$0.2 pixels, or 2 mas.  By centroiding
a map of $\chi^2$, which itself is computed only at integer pixel offsets, our new method avoids interpolating images or PSF
templates.  This makes it relatively fast and free of
systematics.

\section{ADI Reduction}

The goal of an ADI reduction process is to subtract the stellar PSF in
a way that maximizes sensitivity to point sources in the residual
images.  In practice, this means finding an algorithm which produces
Gaussian residuals and an optimal signal-to-noise ratio for single
point sources.

We implement two basic techniques to model the PSF for each frame;
these may be used alone or in conjunction with one another.  The first
method is to take the median of all of the frames to be the model PSF,
subtract this from each individual image, and finally de-rotate and
coadd the sequence.  The second technique is Locally Optimized
Combination of Images (LOCI), described by
\cite{Lafreniere+Marois+Doyon+etal_2007}.  We discuss several
modifications of the basic LOCI algorithm, some of which have been
described elsewhere.  Unfortunately, the AO system on the Subaru
Telescope does not perform well enough in the $H$-band to take
advantage of most of these techniques.  

We characterize each data reduction algorithm by its {\it effective
PSF}, which we define to be the difference between a reduced image
with and without a faint point source.  The effective PSF varies with
the choice of data reduction algorithm and with position on the
detector; it is a product both of hardware (the PSF itself) and of software.  For most SEEDS datasets, we find that the basic LOCI
algorithm offers the best compromise between sensitivity and
simplicity.  When SCExAO \citep{Guyon+Martinache+Clergeon+etal_2011}, the new higher-order AO system for the
Subaru Telescope, is fully operational, other algorithms may offer
sigificant sensitivity improvements.

\subsection{Median PSF Subtraction} \label{sec:medpsf}

A simple way to model the PSF is to use the median of all frames in an
ADI sequence (c.f., \citealt{Marois+Lafreniere+Doyon+etal_2006}).  The model PSF will then include all structures and
companions in the FOV averaged over all position angles in the image
sequence.  Azimuthally extended sources will appear in the model PSF
much as they do in individual exposures, while point sources will be
smeared out by field rotation.  

In this simple technique, the same model PSF is subtracted from each
frame.  The frames are then de-rotated to a common reference position
and co-added.  Variations in the PSF, such as a changing Strehl ratio,
will strongly degrade the sensitivity of the final, processed image to
point sources.  A point source itself will suffer from a fractional
flux loss roughly proportional to the ratio of the size of the PSF
core to the amount of field rotation at its location,
\begin{equation}
f_{\rm loss} \sim \frac{\lambda}{D} \frac{1}{r_{\rm sep} \Delta
  \phi}~,
\label{eq:medsub}
\end{equation}
where $\Delta \phi$ is the total field rotation and $r_{\rm sep}$ is
the angular separation between the point source and the central star.
For a companion at $r_{\rm sep} \sim 1''$ with an H-band PSF core
$\lambda/D \sim 0.\!\!''05$ and a total field rotation of $60^\circ$,
$f_{\rm loss} \sim 5$\%.  

Azimuthally extended sources, like disks, will be suppressed by a factor
\begin{equation}
f_{\rm loss}(\phi) \sim 
\int_{\phi - \Delta \phi/2}^{\phi + \Delta \phi/2}
\frac{I(\theta)\,d\theta}{I(\phi)}~.
\end{equation}
If we expand $I$ as a Taylor series about $\phi$, all of the odd terms
vanish due to the symmetry of the integral.  In other words, azimuthal
gradients, as well as azimuthally symmetric sources, are completely
suppressed; only higher-order features survive a median PSF
subtraction.

The left two panels of Figure \ref{fig:loci_psfs} show
the original PSFs in annuli, smeared out by the field rotation of the
dataset, and the effective PSFs after a mean PSF subtraction.  The
latter are, on average, identical to the effective PSFs produced by a
median PSF subtraction.  The dataset shown had 155 exposures, with a
total field rotation of $\sim$30$^\circ$, and is typical of a SEEDS
observation.  The integrated flux in the mean PSF subtracted image is
zero to within 0.1\% of the flux in the raw PSF image.  

Median PSF subtraction is simple both conceptually and
computationally, and has been successfully used to measure the
geometry of circumstellar disks
\citep{Thalmann+Janson+Buenzli+etal_2011}.  However, as discussed
above, this technique preserves only high-azimuthal-order disk
features; as a result, it is only effective on a small subset of the
SEEDS disk sample.  Furthermore, other techniques such as LOCI,
described in the following section, are much more sensitive to point
sources.  We include median subtraction as an option in ACORNS-ADI but generally recommend against its use.  We use it
here mainly as a baseline against which to measure the performance of
other algorithms.

\subsection{LOCI} \label{sec:loci}

LOCI, a technique for empirical PSF modeling, was introduced by
\cite{Lafreniere+Marois+Doyon+etal_2007}.  The LOCI algorithm models
the PSF in an ADI frame as a local linear combination of other frames
in the sequence, with the coefficients calculated using simple
least-squares.  In each region of frame $i$, LOCI takes the other
frames $\{j\}$ and fits coefficients $\{\alpha_{ij}\}$, eventually
producing an image of residual intensity,
\begin{equation}
{\cal R}_i = I_i - \sum_j \alpha_{ij} I_j~.
\label{eq:basic_loci}
\end{equation}
LOCI fits for the $\{\alpha_{ij}\}$ over optimization regions
typically several hundred PSF footprints---several thousand
pixels---in size.  The subtraction regions are generally at least a factor of ten smaller.  We give more details about the calculation of the LOCI coefficients in
Appendix \ref{app:locisub}.  Because LOCI uses least-squares fitting,
the solution for the $\{\alpha_{ij}\}$ is a linear problem, with the
size of the resulting linear system set by the number of frames in an
ADI sequence.

\begin{figure*}
\begin{center}
\includegraphics[width=0.87\linewidth]{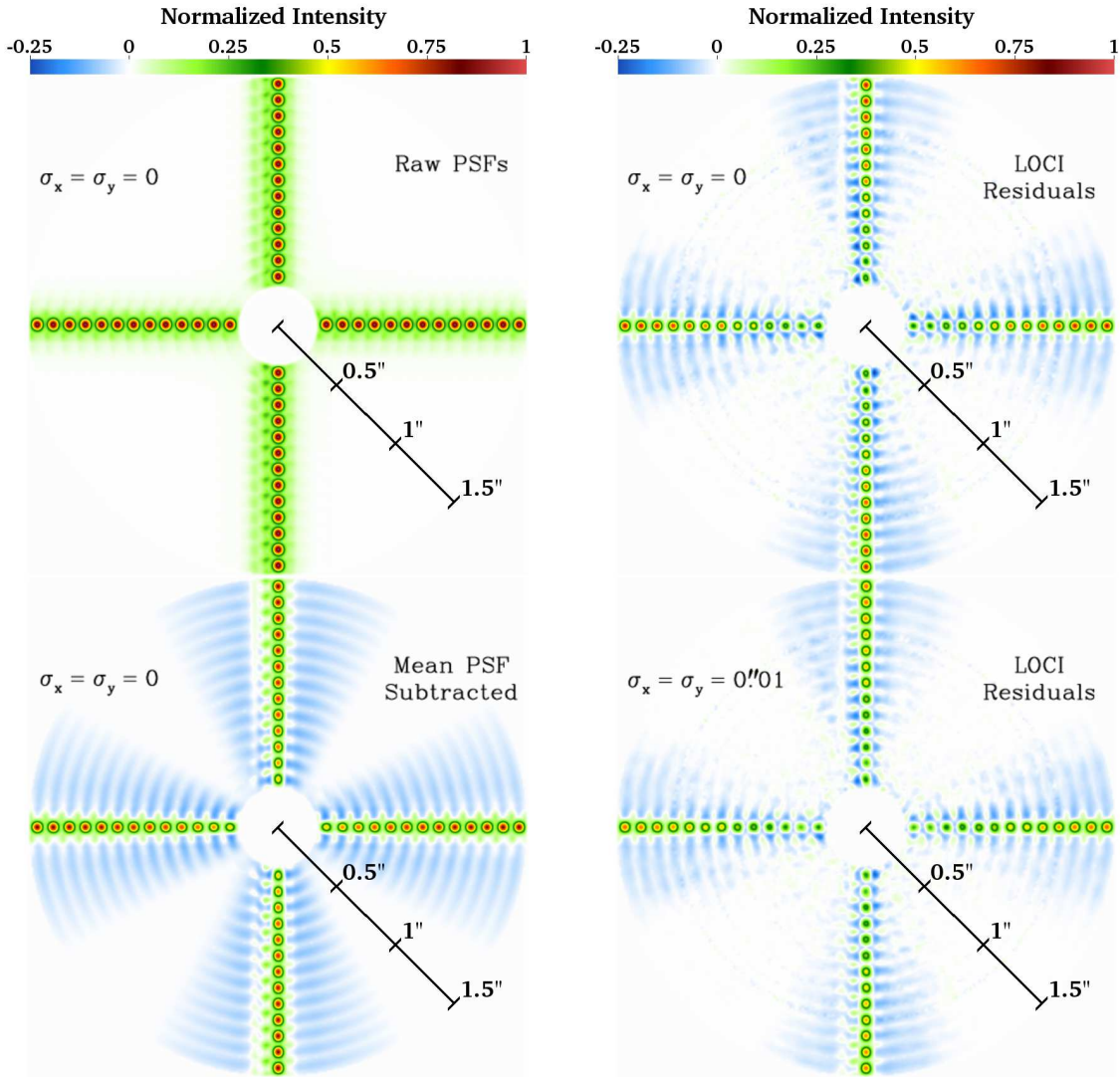}
\end{center}
\caption{The effect of ADI image processing on the original H-band PSF
  (top left panel); all point sources in all panels were reduced
  independently of one another.  We normalize the intensity to the peak in the original PSF.  Lower-left panel: the effect of
  subtracting a mean companion PSF; for a faint source, this is
  equivalent to a median PSF subtraction (Equation \eqref{eq:medsub}).
  The right panels show the residuals after running a LOCI reduction
  \citep{Lafreniere+Marois+Doyon+etal_2007}, with (bottom-right) and without (top-right) errors in imaging registration $\sigma_x$ and $\sigma_y$ prior to the reduction.  A random Gaussian positional error of $0.\!\!''01$, 1/6 of the PSF
  FWHM, reduces the flux in the effective PSF core by $\sim$20\%.  The integrated flux in
  each ADI-processed PSF is approximately zero.  The LOCI-processed
  PSFs are difficult to characterize at small radii, and at all radii,
  they depend strongly on the precision and stability of the image
  registration.}
\label{fig:loci_psfs}
\end{figure*}

To interpret a LOCI-processed ADI sequence, artificial point sources are
added and the data are re-reduced
\citep{Lafreniere+Marois+Doyon+etal_2007}.  Here, we define the LOCI
effective PSF to be the difference between the final, reduced image
with and without a faint point source.  The right panels of Figure
\ref{fig:loci_psfs} show the effective PSF after reducing a sample
SEEDS dataset with LOCI.  The dataset, with 155 frames and a total
field rotation of $\sim$30$^\circ$, represents a typical SEEDS observation.
To ensure that each effective PSF is independent from the others, we
add and reduce the faint companions one at a time.  We use
optimization regions 200 PSF footprints in size and a minimum field
rotation of 70\% of the PSF full width at half maximum (FWHM) as our fiducial LOCI parameters.  

As with median PSF subtraction (Section \ref{sec:medpsf}), LOCI
subtracts azimuthally displaced copies of a faint source, producing
negative ``wings'' in the effective PSF with an integrated flux
approximately equal to the flux in the core.  Indeed, the integrated flux
in each LOCI panel is zero to within 0.5\% of the flux in the original
PSFs (top-left panel).  The bottom-right panel shows
the effect of a random positional jitter on the effective PSFs; such a
jitter could be due to unmodeled instabilities in the PSF or image
registration errors.  These jitters, even if only 1/6 of the PSF FWHM
in each coordinate, smear out the PSF cores and significantly degrade
the sensitivity of observations (see Figure \ref{fig:loci_psfs}), emphasizing the need for reliable,
sub-pixel image registration.

\begin{figure*}
\includegraphics[width=\linewidth]{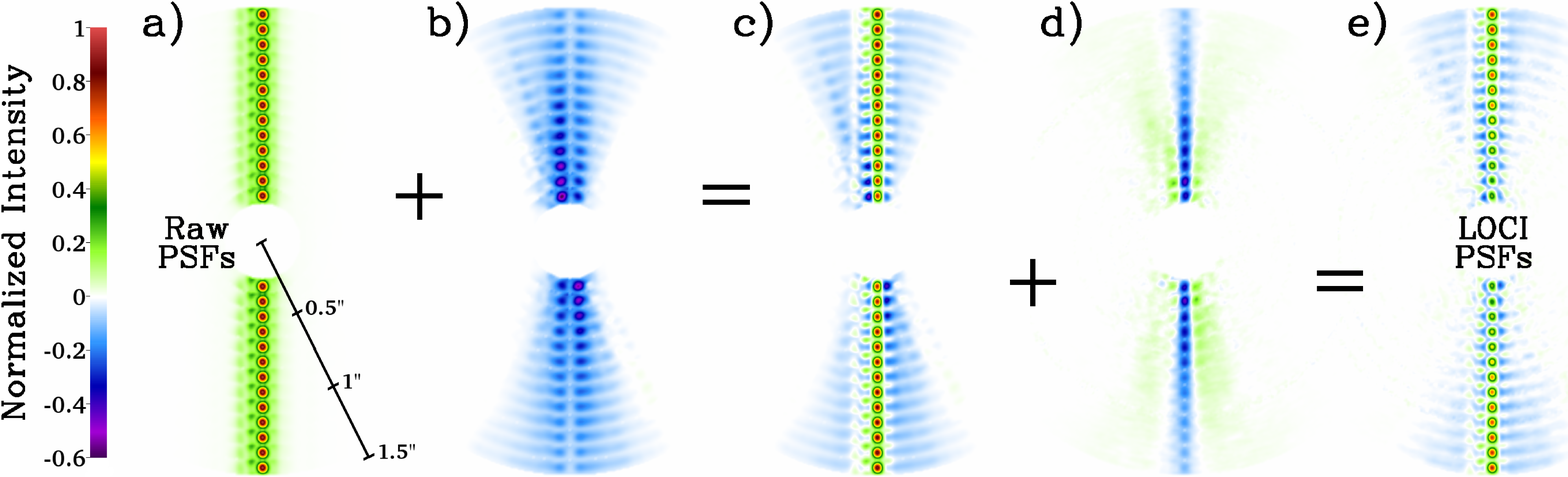}
\caption{A decomposition of LOCI image processing.  Panel a) shows the
  original PSFs, Panel b) shows the effect of subtracting angularly
  displaced copies of the PSF weighted by the LOCI subtraction
  coefficients, and Panel c) is the sum of Panels a) and b).  Panel d)
  presents an additional source of flux suppression, which we
  demonstrate in Appendix \ref{app:locisub}: if we perturb the flux in
  an image, LOCI will, on average, fit this perturbed flux with a
  coefficient between 0 and 1.  In Panel d), this coefficient varies
  from $\sim$0.5 at $0.\!\!''3$ to $\sim$0.1 at $1.\!\!''5$.  We
  measure both effects as a function of spatial position and combine
  them to produce our sensitivity maps.  Panel e), the sum of Panels
  c) and d), shows the effective PSFs (see Figure
  \ref{fig:loci_psfs}).}
\label{fig:partial_sub_process}
\end{figure*}

\subsection{Calibrating LOCI} \label{sec:loci_calib}

Especially at small separations from the central star, LOCI suppresses
companion flux more than does median PSF subtraction.  This is
partly because the best reference frames tend to be nearby in time
(and hence have little relative field rotation).  Panels a)--c) of
Figure \ref{fig:partial_sub_process} demonstrate this effect.  Panel
b) shows azimuthally-displaced PSF copies weighted by the LOCI
coefficients obtained without adding a faint companion; Panel c),
which closely resembles the mean-subtracted (lower left) panel of Figure
\ref{fig:loci_psfs}, shows the effective PSFs after this step.  Because of 
LOCI's angular protection criterion, there is little intensity suppression in the PSF cores in Panel b).  
However, this does not account for the full amount of flux loss (Panel e)).

As we show in Appendix \ref{app:locisub}, the addition of a faint source 
perturbs the LOCI coefficients themselves.  Rather than minimizing the 
least squares equation for the companion-free PSF (Equation \eqref{eq:basic_loci} 
squared and summed over pixels), we instead minimize 
\begin{equation}
\sum_{\rm pixels} \left( (I_i + I'_i) - \sum_j \alpha'_{ij} \left( I_j + I'_j \right) \right)^2~,
\label{eq:loci_perturb}
\end{equation}
where $I'_i$ is the perturbing (companion) intensity in frame $i$ from a faint 
companion, and $\{\alpha'_{ij}\}$ are the perturbed LOCI coefficients.  We use an 
approximation for $I'_i$, minimize Equation \eqref{eq:loci_perturb}, and 
linearize it about the unperturbed coefficients $\{\alpha_{ij}\}$ to derive the fractional flux 
suppression.  Appendix \ref{app:locisub} gives the full derivation.  
This additional effect is the reason why adding too many reference frames can 
actually degrade LOCI's performance.  To a certain degree, which varies according to AO performance and radial
separation (and the number of reference frames), LOCI can fit anything.  Panel d) shows this effect in the
same SEEDS data sequence; at small separations, it can suppress
companion flux by an additional factor of $\sim$2.

\begin{figure}
\includegraphics[width=\linewidth]{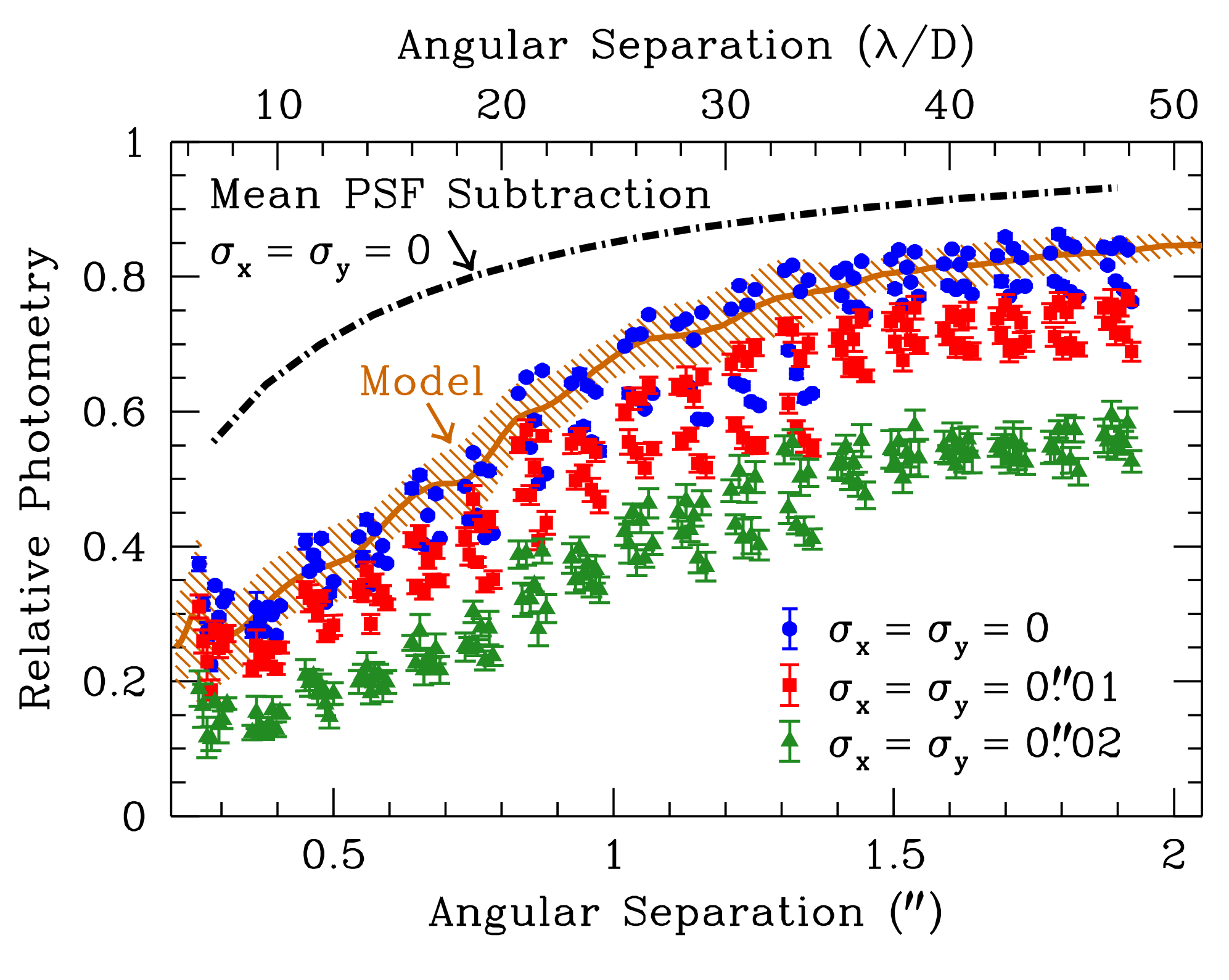}
\caption{Aperture photometry of reduced point sources normalized to
  the photometry of the original PSF.  The error bars are the standard
  deviations of sources of various intensities at fixed position,
  while the vertical scatter of points represents azimuthal variation.
  The effective PSF is linear in source flux (Appendix
  \ref{app:locisub}), but varies by up to $\sim$20\% with
  azimuthal position.  The orange curve shows the radial profile of a
  map of fractional flux suppression computed as described in Appendix
  \ref{app:locisub}, while the hatched region indicates
  $\pm2\sigma$ of azimuthal variation.  Though it does not capture the
  full range of azimuthal variation, our estimated flux suppression
  matches the simulated sources (blue points) with no systematics.
  Fluctuations in the position and shape of the PSF core and image
  registration errors, even with standard deviations ($\sigma_x$, $\sigma_y$)
  that are small compared with the PSF FWHM
  ($0.\!\!''06$), can introduce large systematic errors in the
  recovered photometry.  }
\label{fig:apphot}
\end{figure}

The product of these two effects, the subtraction of azimuthally displaced PSF copies
and the perturbation of the LOCI equations, accounts for the suppression of companion flux in a
LOCI reduction.  Both vary as a function of position but are nearly
independent of companion flux.  Figure \ref{fig:apphot} shows the
results of aperture photometry on actual LOCI PSFs like those in the
right panels of Figure \ref{fig:loci_psfs}.  The error bars on
individual points indicate the scatter in relative photometry as a
function of source flux; that these are zero for the blue points
(those with no positional jitter) indicates that the LOCI effective
PSFs are linear in source flux.  However, positional jitter, whether
from AO tracking errors or from poor image registration, can introduce
significant systematic errors into the recovered sensitivities.

To avoid adding test sources everywhere on the field-of-view, we
produce a map of flux suppression using the method derived in
Appendix \ref{app:locisub}.  As an alternative, we could add faint
sources to densely populate the FOV.  However, {\it these
sources would have to be reduced independently of one another}.  If
there is more than one source in an optimization region, the effect of
the sources on the LOCI subtraction coefficients will change (see
Equation \eqref{eq:partialsubfeedback}).  In general, because the
linear system will be more heavily constrained, the residual intensity
will be larger, and the user will overestimate his or her sensitivity.

The orange curve in Figure \ref{fig:apphot} indicates the radial
profile of our map of simulated flux loss, with the hatched region
covering a spread of $\pm2\sigma$.  Because we do not compute the
perturbation of the LOCI coefficients self-consistently, and because
we neglect asymmetries in the companion PSF, we do not capture the
full range of positional variability in relative photometry.  
For this reason, we recommend using the mean flux suppression at a 
companion's separation to calibrate its flux.  We also recommend adding the azimuthal
variance in the partial flux subtraction to the usual annular variance 
in intensity.  Our model is computationally simple and generally does an excellent
job of reproducing the typical relative photometry at all angular
separations; it is also free of the systematic error that we would
introduce by adding and reducing many point sources simultaneously.

\subsection{LOCI Refinements}

Several authors \citep[e.g.][]{Marois+Macintosh+Veran_2010, Soummer+Hagan+Pueyo+etal_2011, Pueyo+Crepp+Vasisht+etal_2012, Currie+Debes+Rodigas+etal_2012} have recently introduced refinements to the LOCI
algorithm discussed above.  These include reducing the size of
subtraction regions to a single pixel, masking a small area around
each subtraction region, preconditioning the design matrix, and selecting relatively correlated subsets of the full data as reference frames.  We
find that for SEEDS data, none of these steps offer a significant
improvement in sensitivity.  The ineffectiveness of masking an area
around each subtraction region is particularly surprising.  While this
refinement does reduce flux suppression, it does so at the cost of
additional noise.  It appears that, for SEEDS data, LOCI is often as
effective at fitting and removing sources as it is at fitting and removing noise.  
Even the subtraction of a radial profile
from each image, a component of the original LOCI algorithm
\citep{Lafreniere+Marois+Doyon+etal_2007}, does not improve
sensitivity in typical SEEDS data.  

One refinement, suggested by \citet{Marois+Macintosh+Veran_2010} and \citet{Pueyo+Crepp+Vasisht+etal_2012}, does improve the sensitivity of some
SEEDS data.  Because LOCI can fit sources and noise equally well when
given enough reference frames, it is preferable to reduce groups of $\sim$100
frames at a time; a number somewhat smaller than the size (in PSF
footprints) of the optimization regions.  We implement this refinement
by adding the capability to process data in an integer number of
groups of frames, with a single group being equivalent to a normal
LOCI reduction.

We introduce three refinements of our own in addition to those listed
above: 
\begin{enumerate}
\item Performing PCA on an ADI sequence and subtracting the first $n$
components before applying LOCI, similar to the methods suggested by \cite{Soummer+Pueyo+Larkin_2012} and \cite{Amara+Quanz_2012};
\item Including principal components as reference frames in the LOCI process; and 
\item Applying LOCI twice, to
over-correct the residuals in the first application.  
\end{enumerate}
While we expect
these refinements to be useful with a higher Strehl ratio, they seem
to suppress noise and sources equally well in SEEDS data with its
$\sim$30\% Strehl ratio in the $H$-band.
\cite{Soummer+Pueyo+Larkin_2012} and \cite{Amara+Quanz_2012} describe algorithms in
which they use a library of PSF components like those we use for image
registration (Section \ref{sec:center}).  Unfortunately, while these
components are sufficiently good to register SEEDS images, they do not
improve the sensitivity of LOCI.  For SEEDS data, they are not even good enough to perform absolute centroiding to better than $\sim$1 pixel.  
Unlike a space telescope, HiCIAO's
AO system must be re-tuned before each observation.  As a result, the
PSF variation from one observation to the next is generally much
larger than the variation within a single ADI sequence.  Applying an
initial PCA subtraction also makes it much more difficult to
understand the fractional flux loss, and hence the sensitivity to
point sources.  In other words, it makes the flux suppression in Panel d) of Figure \ref{fig:partial_sub_process} significantly larger and harder to estimate.  

We implement all of these refinements as optional
features of ACORNS-ADI.  With the improved
performance of SCExAO, Subaru's next-generation AO
system, or when applied to data from other instruments, these refinements may become much more powerful.

\section{Searching for Point Sources}

After reducing each frame in an ADI sequence, we combine them into
a single image and search for point sources.  We now
discuss each step in turn.  In Section \ref{sec:mean_median}, we introduce a new algorithm,
intermediate between the mean and the median, to combine an image
sequence.  This new algorithm
improves the standard deviation by up to 20\% relative to taking the
median of the images.  In Section \ref{sec:filter}, we test several filters to
search for point sources, settling on a $0.\!\!''05$-diameter circular
aperture as the best choice.

\subsection{Combining an Image Sequence} \label{sec:mean_median}

The optimal method to combine a sequence of $N$ frames ($N \gg 1$) into a final,
reduced image depends on the properties of the errors in each frame.
For example, if the errors are independent and normally distributed,
taking the mean of all of the frames gives a combined datapoint with only $4N / (\pi (2N-1)) \approx 64\%$ of the variance 
obtained by taking their median \citep{Kenney+Keeping_1962}.  However, using the
mean is not robust to outliers, and is a poor choice at small angular
separations where speckle residuals may be highly non-Gaussian between
frames.

The original LOCI algorithm \citep{Lafreniere+Marois+Doyon+etal_2007}
and various refinements \citep[e.g.~][]{Soummer+Hagan+Pueyo+etal_2011} simply
use the median of their LOCI-processed frames, which may not be
optimal for HiCIAO data, particularly in regions far from the central
star where we expect read noise to dominate.  We therefore use the trimmed mean,
which is continuous between the mean and
median: we sort the image sequence at each spatial location, and take
the mean of the middle $n$ points, discarding $(N-n)/2$ values each at the high and low end.  When $n = 1$, this is equivalent
to taking the median; when $n = N$, the number of images in the
sequence, it is equivalent to taking the mean.  We derive the
efficiency of this estimator for data drawn from a normal distribution
in Appendix \ref{app:meanmedstats}.

\begin{figure}
\includegraphics[width=\linewidth]{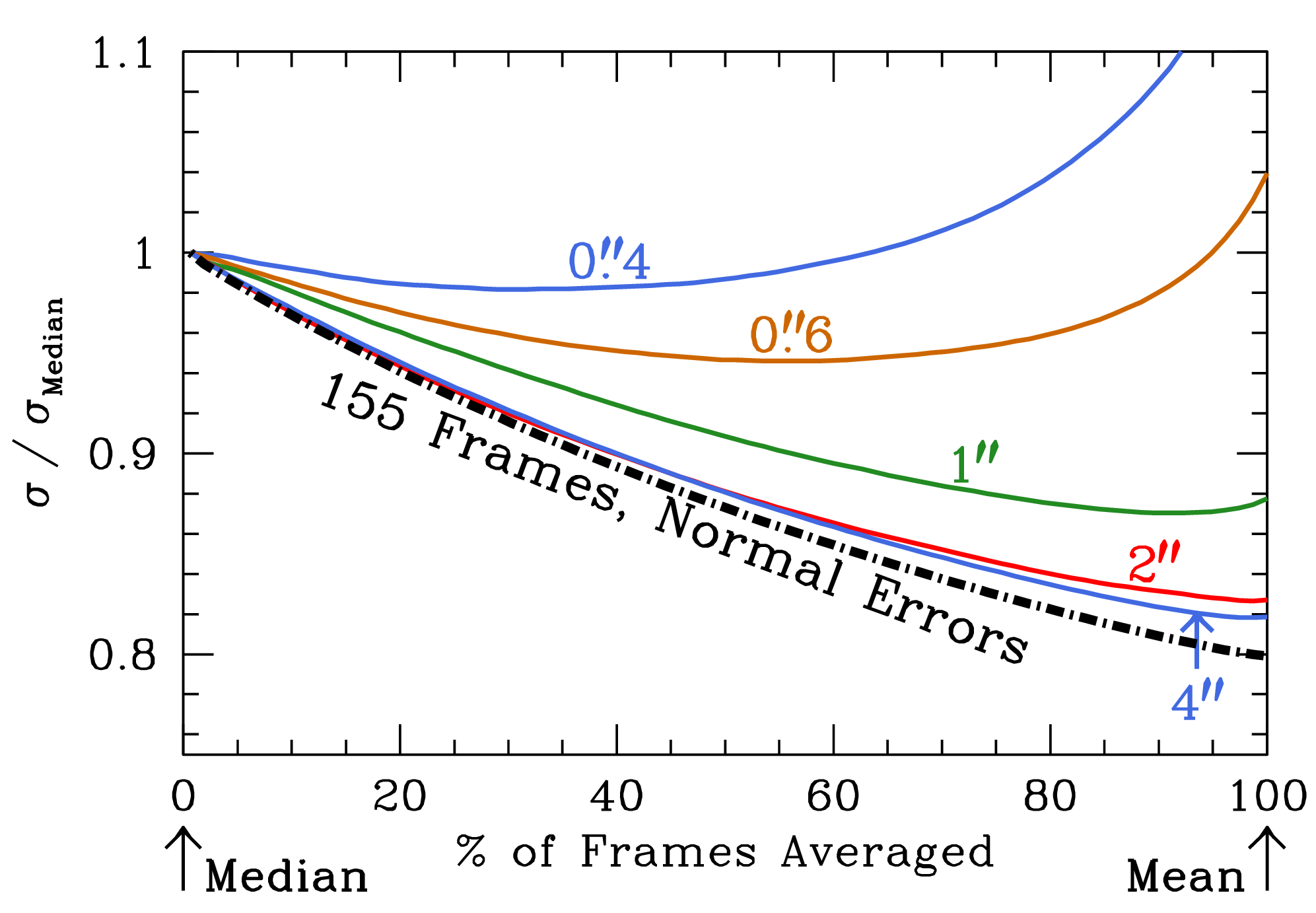}
\includegraphics[width=\linewidth]{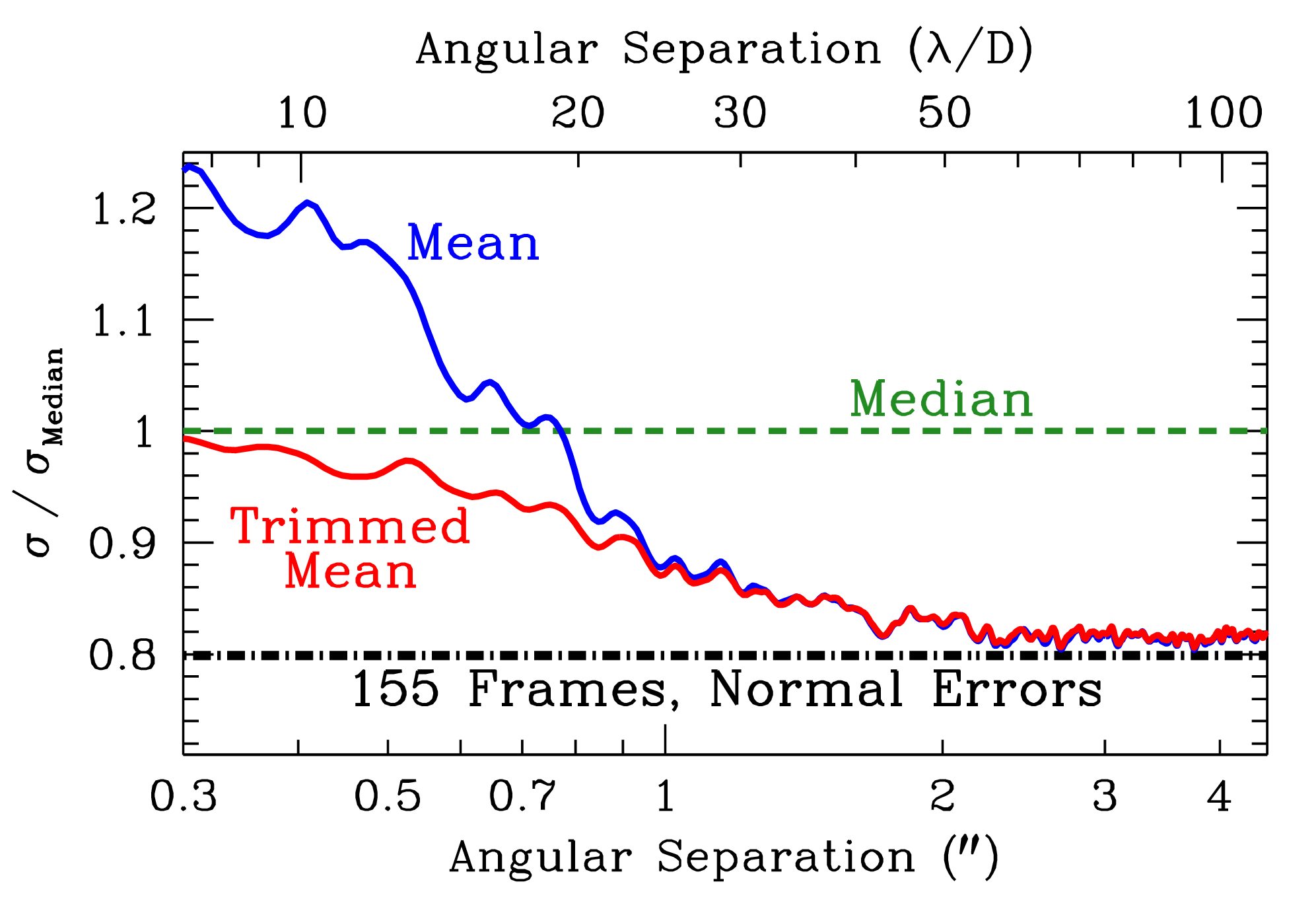}
\caption{Top panel: relative efficiency of the trimmed mean (Section \ref{sec:mean_median}) as a function of the fraction of data averaged.  At small separations, like $0.\!\!''4$ (top blue curve), the median out-performs the mean, while the reverse is true far from the central star (red and bottom blue curves).  However, a trimmed mean, intermediate between the mean and median,
out-performs both.  Bottom panel: the noise profile in a final, reduced image relative to a median combination of all frames (green dashed line).  The blue curve indicated a mean of all of the frames, while the black dot-dashed line indicates the theoretical value ($\approx \sqrt{2/\pi}$) for Gaussian data.  The trimmed mean (red curve) out-performs the median at all radii, and by nearly 20\% far from the central star.}
\label{fig:mean_median}
\end{figure}

The top panel of Figure \ref{fig:mean_median} shows this estimator as
applied to a sample HiCIAO image sequence of 155 frames.  At small
angular separations, outliers are relatively common and the mean
provides a poor estimator, with a standard deviation $\sim$20\% higher
than that of the median.  Far from the central star, however, the
picture is reversed; using the mean gives a large improvement in
sensitivity.  At nearly all separations, the optimal solution is
somewhere in between, close to the median at small separations and
close to the mean further away.  We expect (and Figure
\ref{fig:mean_median} confirms) that the relationship between the
optimal $n$ in the trimmed mean and angular separation from the
central star is essentially monotonic.

We implement a trimmed mean estimator iteratively.  We begin with the
median of our image sequence and calculate its noise profile.  We then
calculate the noise profile for an image created by averaging more
frames (or trimming fewer), and replace data points in the median image at annuli where
the new estimator reduces the variance.  We repeat this step, using
more frames (larger $n$) in each successive estimator, until we use
nearly all the frames.  We always trim at least 5\% of the data to
guard against cosmic rays and rare outliers; Figure
\ref{fig:mean_median} shows that this approximation incurs at most
a 0.5\% penalty in noise.  

The bottom panel of Figure \ref{fig:mean_median} shows the results of our iterative
trimmed mean relative to a simple median of the frames in an image
sequence.  The new image represents an improvement in noise at all
angular separations, with a $\sim$20\% improvement at large
separations where the frame-to-frame noise is very nearly Gaussian.  

While the frame-to-frame noise at a given pixel may be significantly non-Gaussian, the distribution of trimmed means, due to the central limit theorem, {\it is} Gaussian.  
Our new final image thus retains the noise
properties of the original LOCI algorithm; in our sample dataset, it produced
zero single-pixel false positives (pixels with $>$$5\sigma$
fluctuations).

\subsection{Filtering the Image} \label{sec:filter}

\begin{figure}
\includegraphics[width=\linewidth]{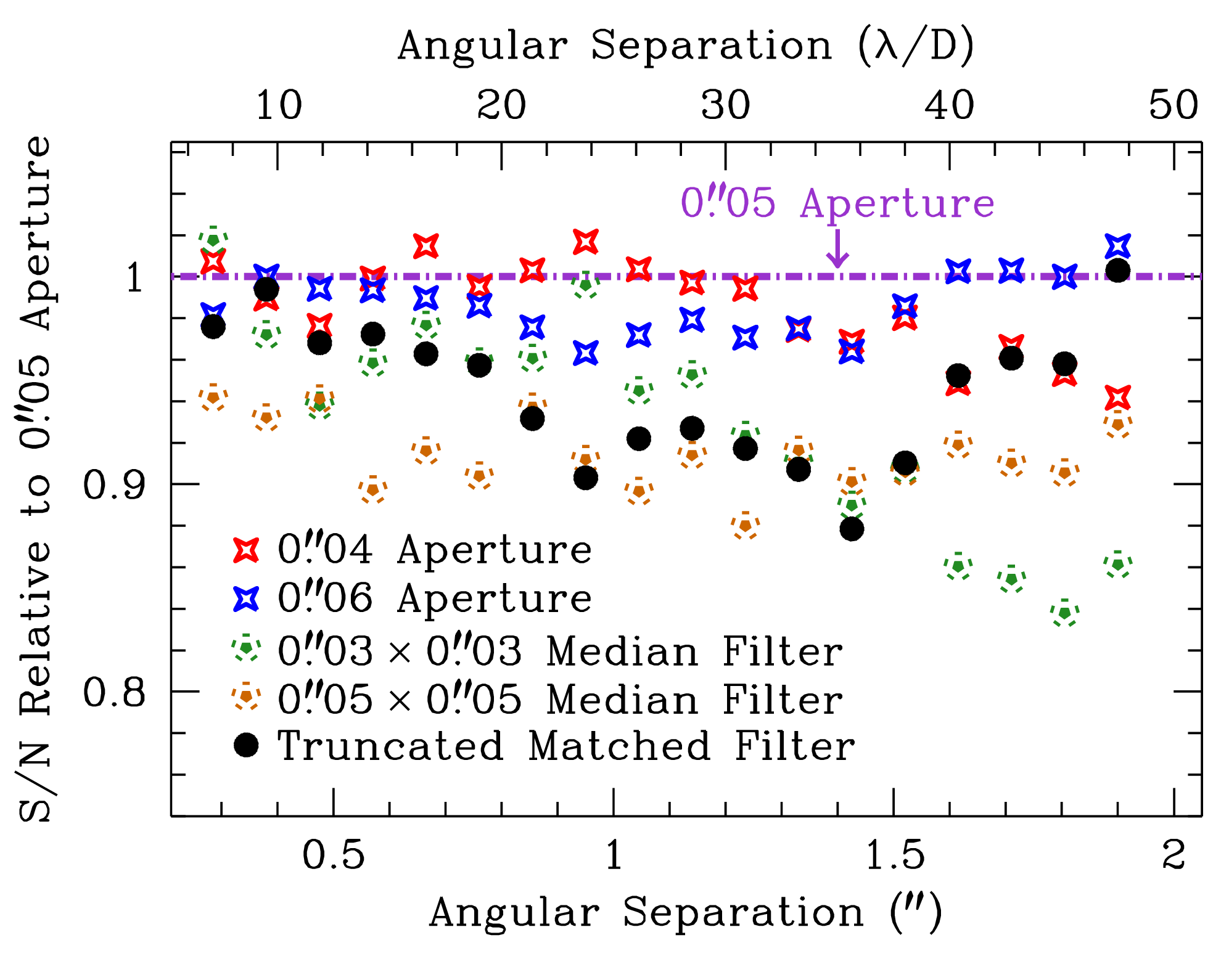}
\caption{Comparison of three detection/photometry algorithms for a
  companion at 1$''$; the raw S/N is the peak companion intensity over
  the noise at the companion's separation.  The relative S/N is
  normalized to the raw S/N.  Aperture photometry is the simplest technique, and generally out-performs median filtering and PSF photometry.  All three methods are linear
  to $\sim$1\% over a wide range of companion fluxes.}
\label{fig:filters}
\end{figure}

The optimal filter for detecting an object depends on both the object
and the character of the noise.  For noise that is Gaussian and 
independent at adjacent pixels, the optimal filter to search for point sources
is the normalized PSF, referred to as a matched filter.  In this
section we measure the performance of three filters for point sources
in the HiCIAO data: 
\begin{enumerate}
\item A matched filter;
\item A circular aperture;
\item A ``truncated'' matched filter set to zero outside an aperture; and 
\item A 2-dimensional median filter.
\end{enumerate}
Figure \ref{fig:filters} shows the relative performance of these
filters.  We truncate the matched filter at twice the radius of our
fiducial $0.\!''05$-diameter aperture to limit the impact of the outer
wings, which can depend strongly on azimuthal position (see Figure
\ref{fig:loci_psfs}).  We do not show the full matched filter, which
fails to out-perform aperture photometry even assuming perfect
knowledge of the effective PSF.  

\begin{figure}
\includegraphics[width=\linewidth]{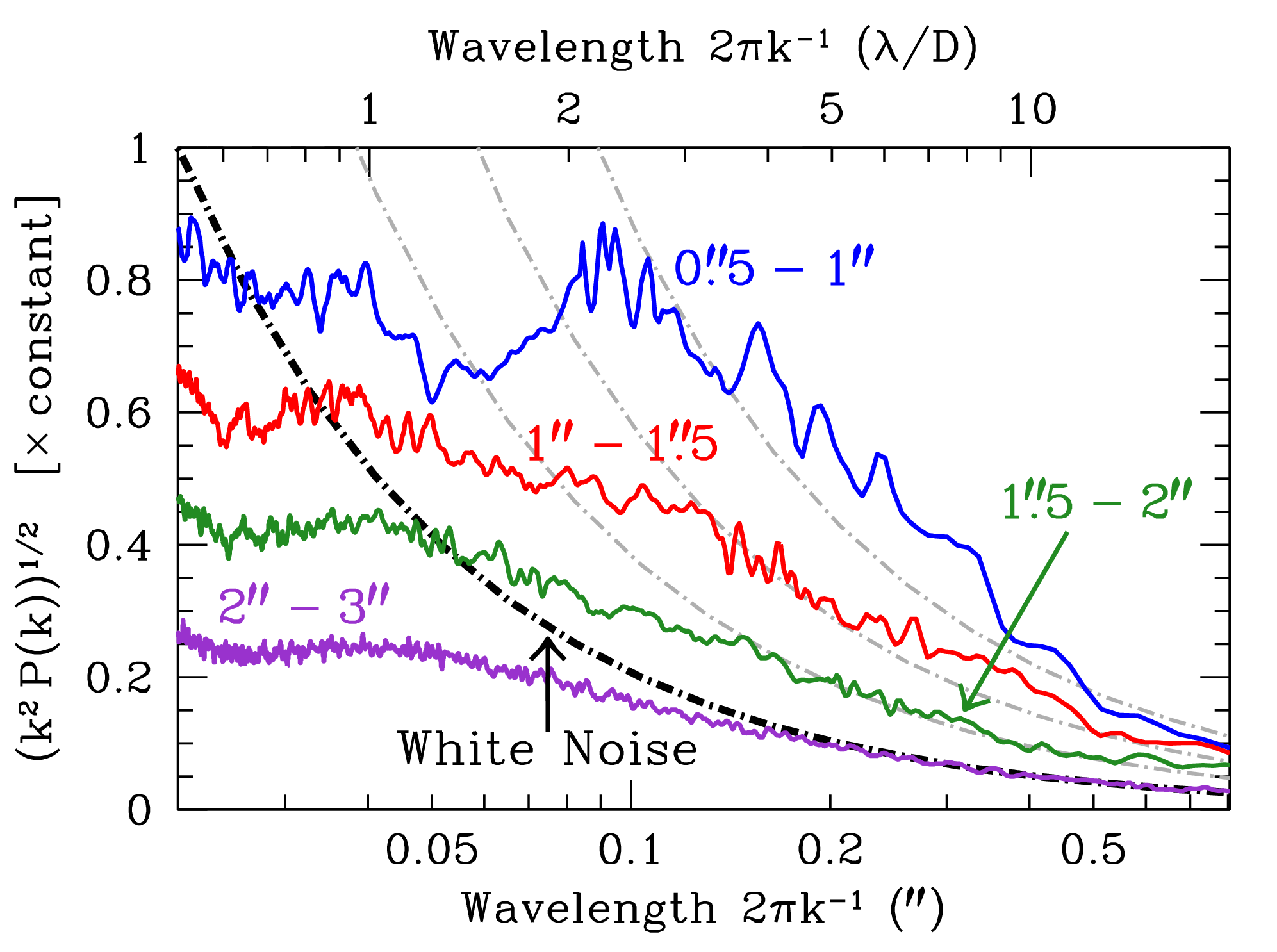}
\caption{Power spectrum of noise at four annuli in our final reduced
  image, with white noise overplotted in black/grey.  The suppression
  of noise at small spatial scales results from averaging.  We
  interpolate each frame onto a new spatial array three times: during
  the distortion correction, for recentering, and finally when we
  derotate each image to a common orientation on-sky.  The increase in
  power at a few $\lambda/D$ is an artifact of LOCI.  Because LOCI
  gives zero average intensity even on modest spatial scales, intensity
  near the central star is anti-correlated over spatial scales larger
  than the PSF.}
\label{fig:noisepowerspec}
\end{figure}

Perhaps surprisingly, a $0.\!''05$-diameter ($1.2 \lambda/D$ at 1.6
$\mu$m) circular aperture seems to offer the best performance of all
the filters.  A two-dimensional median filter is not optimal,
especially far from the central star, because the noise in the reduced
frame is approximately Gaussian.  The poor performance of the matched
filter, on the other hand, results from the strong correlation of the
residual intensity in neighboring pixels.  Figure
\ref{fig:noisepowerspec} demonstrates this correlation in Fourier
space; it results from averaging adjacent pixels when interpolating
onto a new spatial array.  We interpolate each frame three times
during the data reduction process: when applying the distortion
correction, when recentering, and finally when derotating each image
to a common orientation on-sky.  We have verified that we can closely
reproduce the power spectrum of noise at large separations by
smoothing white noise.  

In addition to the suppression of noise at high spatial frequency,
Figure \ref{fig:noisepowerspec} shows an increase in power at a few
$\lambda/D$ (a few PSF diameters), particularly in regions close to
the central star.  This is an artifact of the LOCI algorithm, which
tends to give zero flux averaged over a spatial region larger than a PSF core.  
As a result, LOCI introduces an
anti-correlation in intensity over scales larger than the size of the
PSF.  This also helps explain the poor performance of the full matched
filter, which would otherwise take advantage of the negative wings in
the effective PSF: large random fluctuations will also tend to be
surrounded by regions of negative intensity.

\subsection{Producing a Sensitivity Map} \label{sec:sensitivity}

To date, the vast majority of high-contrast direct imaging
observations have not detected substellar companions.  However, non-detections may still be used to test models of planet and brown dwarf frequency, separation,
and luminosity (e.g., \citealt{Lafreniere+Marois+Doyon+etal_2007}, \citealt{Bonavita+Chauvin+Desidera+etal_2012}).  Such analyses rely on the accuracy of
sensitivity maps.  As we show in Figure \ref{fig:apphot}, it is easy
to systematically overestimate sensitivity by failing to include
effects such as PSF fluctuations and image registration errors.

It has become widely accepted within the community to estimate sensitivity by computing the standard deviation in the final, combined image in annuli around the central star (e.g., \citealt{Lafreniere+Marois+Doyon+etal_2007, McElwain+Larkin+Metchev+etal_2008, Metchev+Hillenbrand_2009, Vigan+Patience+Marois+etal_2012}).  At separations of more than a few $\lambda/D$, these annuli are already at least several tens of PSF footprints in size.  To account for LOCI's suppression of companion flux, the dataset is re-reduced after adding faint sources to compute the fractional flux loss as a function of radial separation.  We begin in the same way, computing the standard deviation of the residual intensity after convolving with our chosen $0.\!\!''05$ aperture.  We correct for companion flux suppression using our own method, which we describe in Section \ref{sec:loci_calib} and Appendix \ref{app:locisub}.  

The result of our sensitivity analysis is not simply a radial profile, but a full 2D sensitivity map, obtained with modest additional computational cost.  As a final step, we multiply this map by the scaled aperture photometry of the central star in a
sequence of reference images, producing a contrast map.

\section{Use and Performance} \label{sec:performance}

ACORNS-ADI is easy to use, and requires user
interaction at only two points: 
\begin{enumerate}
\item To start the program, select the data and calibration files, and
  the reduction parameters; and
\item To interactively set and verify the absolute image centroid.  
\end{enumerate}
The total human time to perform a reduction is thus a couple of
minutes, and is independent of the dataset.  

The amount of computer time required scales with the size of the
dataset and the number of processors available.  ACORNS-ADI is
efficiently parallelized, running more than 12 times as fast on 16 processors as
on a single processor.  A full reduction from raw data on an ADI
sequence of 155 frames takes about 40 minutes on our three-year-old
16-core machine; this compares with $\sim$2 hours of computer time and 8
user interactions to process 1/3 as many optimization regions with serial IDL software adapted from \cite{Lafreniere+Marois+Doyon+etal_2007}.  The LOCI step
scales differently depending on the number of frames.  For datasets of
$\gtrsim$100 frames, it scales with $N_{\rm frames}^4$, while for much
smaller datasets it scales with $N_{\rm frames}^2$.  The other steps
in the data reduction process, with the exception of combining the
images (which is computationally cheap), each scale linearly with
$N_{\rm frames}$.

\subsection{Extension to Other Instruments}

Most of the algorithms presented above apply to ADI data taken by any instrument.  ACORNS-ADI can reduce these data with minimal modification by the user, who must supply:
\begin{enumerate}
\item A flat-field correction and hot pixel mask;
\item An (optional) field distortion correction;
\item A set of PSF templates; and 
\item A curve describing the integrated overlap of PSF cores.
\end{enumerate}
The most difficult item to compute is the set of PSF templates.  We hope to supply a set of images for each of several instruments over the coming months.  The last item simply refers to the flux in an aperture displaced a certain number of pixels from the PSF centroid, and is used to compute the fractional flux loss in LOCI.

While it is easy to extend ACORNS-ADI to other instruments, we offer several notes of caution when doing so.  Different instruments have different conventions for such header data as the exposure time, the number of coadds in a frame, and the image orientation with the image rotator off.  For example, in HiCIAO, the keyword \verb|'EXPTIME'| refers to the total integration time of the frame, while for NIRI, \verb|'EXPTIME'| refers to the integration time for each coadd.  This makes it difficult to write a fully general software package.  Unfortunately, ACORNS-ADI cannot detect these differing conventions automatically, and the user must be cautious.

\section{Conclusions}

We have described ACORNS-ADI, a new, parallel, open-source software package for
reducing ADI data from the SEEDS survey.  Most of its modules apply
equally well to non-SEEDS ADI data, and the entire package could
easily be adapted to analyze data from other instruments.  We have
introduced three new algorithms: 
\begin{enumerate}
\item A new method of performing image registration, which is accurate
  to $\sim$$0.\!\!''002$, $\sim$0.2 HiCIAO pixels;
\item A new method for combining images in an ADI sequence, which
  reduces noise by up to 20\%; and
\item A new method for calculating the flux loss in the LOCI algorithm
  without adding and reducing artificial point sources.  
\end{enumerate}
These new algorithms may be applied to any ADI dataset, improving
performance and decreasing run time.

We have described and characterized each step of the ADI data
reduction process for SEEDS data.  With ACORNS-ADI, we will be
able to process data much more quickly and efficiently, taking
advantage of the SEEDS survey's design as a large strategic observing
program.  In the future, we will modify ACORNS-ADI to process data
from other surveys and instruments, providing a large set of uniformly
reduced data with which to perform statistical analyses of substellar
companion frequencies and luminosities.  

\acknowledgments{The authors thank the anonymous referee for many helpful 
comments and suggestions that clarified this manuscript.
This research is based on data collected at the Subaru 
Telescope, which is operated by the National Astronomical Observatories of Japan.
This material is based upon work supported by the National Science
  Foundation Graduate Research Fellowship under Grant No.~DGE-0646086.  
Part of this research was carried out at the Jet Propulsion Laboratory,
California Institute of Technology, under a contract wiht the National
Aeronautics and Space Administration.  The authors wish to recognize and acknowledge the very significant cultural role and reverence that the summit of Mauna Kea has always had  within the indigenous Hawaiian community.  We are most fortunate to have the opportunity to conduct observations from this mountain. }

\appendix

\section{A: Partial Subtraction in LOCI} \label{app:locisub}

In LOCI, we build a model PSF for each frame $I_i$ in an ADI sequence
from the other frames $\{I_j\}$ satisfying LOCI's angular displacement
criterion.  Denoting the intensity at pixel $k$ in frame $j$ by
$I_{jk}$, we calculate the coefficients $\{ \alpha_j \}$ that minimize
\begin{equation}
R_i^2 = \sum_{{\rm pixels}~k} \left( I_{ik} - \sum_{{\rm frames}~j} \alpha_j I_{jk} \right)^2~,
\label{eq:loci_min}
\end{equation}
where the first sum is over pixels $k$ in the optimization region.
The coefficients $\{ \alpha_j \}$ are the solution to the linear
system
\begin{equation}
\mathbf{A} \cdot \boldsymbol{\alpha} = \mathbf{b}~,
\label{eq:alpha}
\end{equation}
with 
\begin{equation}
{\rm A}_{jl} = \sum_{{\rm pixels}~k} I_{jk} I_{lk} \quad {\rm and} \quad
{\rm b}_j = \sum_{{\rm pixels}~k} I_{ik} I_{jk}~.
\end{equation}
We can perturb this problem by adding a faint source of intensity $I'$
to frame $i$, and azimuthally displaced copies of it to the other
frames.  We approximate the azimuthally displaced copies by
adding a source of intensity
\begin{equation}
I'_{\rm eff} = I'-\sum_j \alpha_j I'(\delta \phi_j)
\label{eq:ieff}
\end{equation}
to frame $i$.  We then solve the perturbed problem by
minimizing Equation \eqref{eq:loci_min} again, this time with the faint 
effective source of Equation \eqref{eq:ieff} added to each frame.
Note that we use the unperturbed coefficients $\{\alpha_j\}$ in Equation 
\eqref{eq:ieff} rather than iteratively solving for the exact perturbed solution.  
The perturbations $\{ \beta_j \}$ in the LOCI
coefficients will then be given by the solution to the linear system
\begin{equation}
\mathbf{A} \cdot \boldsymbol{\beta} = \mathbf{b'}~,
\label{eq:beta}
\end{equation}
with 
\begin{equation}
{\rm b}'_j = \sum_{{\rm pixels}~k} 
\left( I'_{k} - \sum_{{\rm frames}~l} \alpha_l I'_{k}(\delta \phi_l) \right)
I_{k}(\delta \phi_j)~.
\label{eq:partialsubfeedback}
\end{equation}
Note that the companion intensity $I'$ and the coefficients $\beta_j$
are both linear in the companion flux.  The residual intensity in
pixel $k$ of frame $i$, $\mathcal{R}_{ik}$, is then
\begin{align}
\mathcal{R}_{ik} &= I_{ik} + I'_{ik} - 
\sum_j \left( \alpha_j I_{jk} + \alpha_j I'_k (\delta \phi_j) + 
\beta_j I_{jk} + \beta_j I'_k (\delta \phi_j) \right) \\
&\approx I_{ik} - \sum_j \alpha_j I_{jk} 
+ I'_{ik} - \sum_j \left(\alpha_j I'_k (\delta \phi_j) + \beta_j I_{jk} \right)~.
\label{eq:loci_linear}
\end{align}
Because the source is faint, we drop the quadratic term $\beta_j
I'_k (\delta \phi_j)$.  The first two terms in Equation \eqref{eq:loci_linear} give
the residual intensity without the additional faint source, while the
latter three terms give the residual intensity in the LOCI-processed
image.  These are all proportional to $I'$ ; hence, the LOCI effective
PSFs are linear in the source flux.

We compute the relative photometry of a LOCI effective PSF by
evaluating the latter two terms of Equation \eqref{eq:loci_linear},
multiplying by an aperture and summing over pixels (as described in
Section \ref{sec:filter}, we use aperture photometry for the SEEDS
data).  We use a source of unit flux,
\begin{equation}
\sum_{{\rm pixels}~k} I'_{ik} a_{k} = 1~,
\end{equation}
where $a$ is the aperture.  Thus, $\sum_k I'_k (\delta \phi_j)$ is the flux in an
aperture displaced from the PSF center by the relative field rotation
between frames $i$ and $j$.  We pre-compute these fluxes as a function of position,
then multiply by the LOCI coefficients $\{\alpha_j\}$.

We estimate the last term in Equation \eqref{eq:loci_linear} by first
approximating $I'_{\rm eff}$ (Equation \eqref{eq:ieff}) as a Gaussian
central peak, with two pairs of wider, negative Gaussians representing the
wings (c.f.~Figure \ref{fig:loci_psfs}).  The angular separations of the
negative Gaussians from the central peak are 1.5 and 3 times LOCI's angular
protection zone, or 1 and 2 times the standard deviation in parallatic angle, 
whichever are less.  Our approximation for $I'_{\rm eff}$ has zero integrated flux,
and accurately recovers the LOCI flux suppression (see Figure \ref{fig:apphot}).  By using the same approximate
$I'_{\rm eff}$ for each frame, we avoid the need to recompute {\bf b$'$}
(Equation \eqref{eq:partialsubfeedback}) for each frame, and save a factor of
nearly $N_{\rm frames}$ in execution time.  We then compute $\{\beta_j\}$
using Equation \eqref{eq:beta}.  Because we have already performed LU
decomposition on $\mathbf{A}$ to solve Equation \eqref{eq:alpha}, this
step takes little computational effort.  We finally compute the total
flux loss within the aperture,
\begin{equation}
\sum_{{\rm pixels}~k} a_k \sum_{{\rm frames}~j} 
\left( \alpha_j I'_k (\delta \phi_j) + \beta_j I_{jk} \right)~.
\label{eq:fracfluxloss}
\end{equation}
This allows us to compute the fractional flux loss everywhere on the
FOV for little additional cost relative to LOCI itself.

\section{B: Efficiency of the Trimmed Mean Estimator} \label{app:meanmedstats}

We briefly derive the efficiency of the trimmed mean
estimator used in Section \ref{sec:mean_median} when applied to data
with normal errors.  
The efficiency is inverse of the variance of an estimator relative to the minimum possible variance of any estimator.  For Gaussian data, the mean provides the minimum possible variance, which is equal to $\sigma^2/N$.
In the trimmed mean, we take the mean of the
middle $n$ out of a total of $N$ points; for simplicity, we
will assume $N$ and $n$ to be odd.  We will work
mostly in the space of the cumulative distribution function (CDF), in which
each realization of the distribution is drawn from a uniform
distribution between 0 and 1.

As a first step, we write down the probability that the middle $n$
realizations (in CDF space) are all between $x_1$ and $x_2$, with one
realization each at $x_1$ and $x_2$ (within $dx$); that is, that we
have $(N-n)/2 - 1$ realizations $<x_1$ and $(N-n)/2 - 1$ realizations
$>x_2$.  We denote $(N-n)/2 - 1$, the number of datapoints trimmed at each end, by $q$.  We have
\begin{align}
p(x_1, x_2) dx^2 &= x_1^q (1 - x_2)^q (x_2 - x_1)^n \times {}_NC_q \times {}_{N-q}C_q \times (n + 2) \times (n + 1) \\
 &= x_1^q (1 - x_2)^q (x_2 - x_1)^n \frac{N!}{(q!)^2 n!}~.
\label{eq:px1_x2}
\end{align}

Now, given $x_1$ and $x_2$, we wish to calculate the variance of the mean of $n$ realizations of the truncated normal
distribution.  We will assume
without loss of generality that the normal distribution has zero mean.
We denote these limits as ${\rm CDF}^{-1}(x_1) = \sigma y_1$
and ${\rm CDF}^{-1}(x_2) = \sigma y_2$, where ${\rm CDF}^{-1}$ is the quartile function and $y_2>y_1$ are both drawn from a normal distribution with unit variance.  
Assuming the $n$ realizations to be independent,
the expectation value of the square of their mean is
\begin{equation}
\sum_{y_1,y_2} p(y_1,y_2) \left[ \frac{n}{x_2 - x_1} 
\int_{\sigma y_1}^{\sigma y_2} \frac{t^2\,dt}{n^2 \sqrt{2\pi\sigma^2}} e^{-t^2/2\sigma^2}
+ \frac{n (n - 1)}{(x_2 - x_1)^2} \left( \int_{\sigma y_1}^{\sigma y_2} \frac{t\,dt}{n \sqrt{2\pi\sigma^2}} 
e^{-t^2/2\sigma^2} \right)^2 \right] ~.
\label{eq:meanmed_var}
\end{equation}
We integrate the first term by parts and then integrate the second term.  The inverse of the efficiency is then
$\sigma^2/N$ times Equation \eqref{eq:meanmed_var}, equal to
\begin{align}
& \frac{N}{n} \sum_{y_1,y_2} \frac{p(y_1,y_2)}{x_2 - x_1} \left[ \frac{y_1}{\sqrt{2\pi}} e^{-y_1^2/2}
- \frac{y_2}{\sqrt{2\pi}} e^{-y_2^2/2} 
+ \int_{y_1}^{y_2} \frac{dt}{\sqrt{2\pi}} e^{-t^2/2} 
+ \frac{n - 1}{2\pi(x_2 - x_1)} \left(e^{-y_1^2/2} - e^{-y_2^2/2} \right)^2 \right] \\
=\,\,& \frac{N}{n} \int_0^1 dx_1 \int_{x_1}^1 dx_2\, \frac{p(x_1,x_2)}{x_2 - x_1} 
\left[ \frac{y_1}{\sqrt{2\pi}} e^{-y_1^2/2}
- \frac{y_2}{\sqrt{2\pi}} e^{-y_2^2/2} + x_2 - x_1
+ \frac{n - 1}{2\pi (x_2 - x_1)} \left(e^{-y_1^2/2} - e^{-y_2^2/2} \right)^2 \right] ~. 
\label{eq:mean_med_integral}
\end{align}
We then substitute for $p(x_1, x_2)$ using Equation \eqref{eq:px1_x2} and evaluate the integral.  In the limit of the median ($n=1$), Equation \eqref{eq:mean_med_integral} reduces to $\pi (2n-1) / 4n$, for an asymptotic efficiency of $2/\pi \approx 64\%$ \citep{Kenney+Keeping_1962}.

\bibliographystyle{apj_eprint}
\bibliography{seeds_refs}

\end{document}